\documentclass
[aps,prl,twocolumn,floats,epsfig,showpacs,superscriptaddress]{revtex4-1}%
\usepackage{graphicx}
\usepackage{dcolumn}
\usepackage{amssymb}
\usepackage{amsmath}
\usepackage{amsfonts}
\usepackage{amssymb}
\usepackage{hyperref}
\usepackage{color}
\usepackage{verbatim}
\usepackage{ifpdf}
\usepackage{xcolor}
\usepackage{xcolor}%
\definecolor{red}{rgb}{0.7,0,0}
\definecolor{green}{rgb}{0.,0.35,0.}
\definecolor{blue}{rgb}{0.2,0.2,0.7}
\definecolor{black}{rgb}{0.15,0.15,.15}

\newcommand{\ad}{a^{\dagger}}

\makeindex
\begin{document}

\title{Majorana edge states 
in two atomic wires coupled by pair-hopping}
\date{\today}

\author{Christina V. Kraus}
\affiliation{Institute for Quantum Optics and Quantum Information of the Austrian Academy of Sciences, A-6020 Innsbruck, Austria }
\affiliation{Institute for Theoretical Physics, Innsbruck University, A-6020 Innsbruck, Austria}
\author{Marcello Dalmonte}
\affiliation{Institute for Quantum Optics and Quantum Information of the Austrian Academy of Sciences, A-6020 Innsbruck, Austria }
\author{Mikhail A. Baranov}
\affiliation{Institute for Quantum Optics and Quantum Information of the Austrian Academy of Sciences, A-6020 Innsbruck, Austria }
\affiliation{Institute for Theoretical Physics, Innsbruck University, A-6020 Innsbruck, Austria}
\affiliation{RRC "Kurchatov Institute", Kurchatov Square 1, 123182, Moscow, Russia}
\author{Andreas M. L\"auchli}
\affiliation{Institute for Theoretical Physics, Innsbruck University, A-6020 Innsbruck, Austria}
\author{P. Zoller}
\affiliation{Institute for Quantum Optics and Quantum Information of the Austrian Academy of Sciences, A-6020 Innsbruck, Austria }
\affiliation{Institute for Theoretical Physics, Innsbruck University, A-6020 Innsbruck, Austria}


\begin{abstract}
We present evidence for the existence of Majorana edge states in a number
conserving theory describing a system of spinless fermions on two wires that
are coupled by a pair hopping. Our analysis is based on the combination of a
qualitative low energy approach and numerical techniques using the Density
Matrix Renormalization Group. We also discuss an experimental realization of
pair-hopping interactions in cold atom gases confined in optical lattices, 
and its possible alternative applications to quantum simulation.

\end{abstract}

\maketitle

At present there is significant interest in identifying physical setups where
Majorana fermions (MFs) \cite{majorana1937} emerge as a collective phenomenon
in many-body quantum systems \cite{wilczek2009}. The motivation behind this
search is two-fold: First, the existence of MFs is intimately linked to the
concept of topological phases and their exploration. Second, MFs provide due
to their topological nature a promising platform for topological quantum
computing and quantum memory \cite{nayak2008, alicea2012, beenakker2012}. In a
seminal paper Kitaev pointed out a route towards the realization of MFs in a
simple many-body system \cite{kitaev2001}: A 1D wire of spinless fermions with
a $p$-wave pairing can exhibit a topologically ordered phase with zero-energy
Majorana edge modes. The key ingredient here is the coupling of the wire to a
superconducting reservoir in a grand canonical setting, which is induced in
complex solid state structures via the so called proximity effect. Building on
this result, a remarkable theoretical and experimental effort has been devoted
in search of alternative settings supporting topological superconductivity in
1D condensed matter systems, such as the combination of spin-orbit coupling,
magnetic fields and s-wave interactions \cite{tezuka2012, lutchyn2010,
oreg2010,tsvelik2011, sau2010, Kouwenhoven, Deng, Das,
Rokhinson,stoudenmire2011}. Alternatively, Majorana physics can be observed
with 1D quantum gases coupled to a particle reservoir represented by molecular
condensates, taking advantage of the unique tools for control and measurements
in atomic systems~\cite{liang2011, diehl2011, nascimbene2012}.
\begin{figure}[t]
\includegraphics[width=1.0\columnwidth]{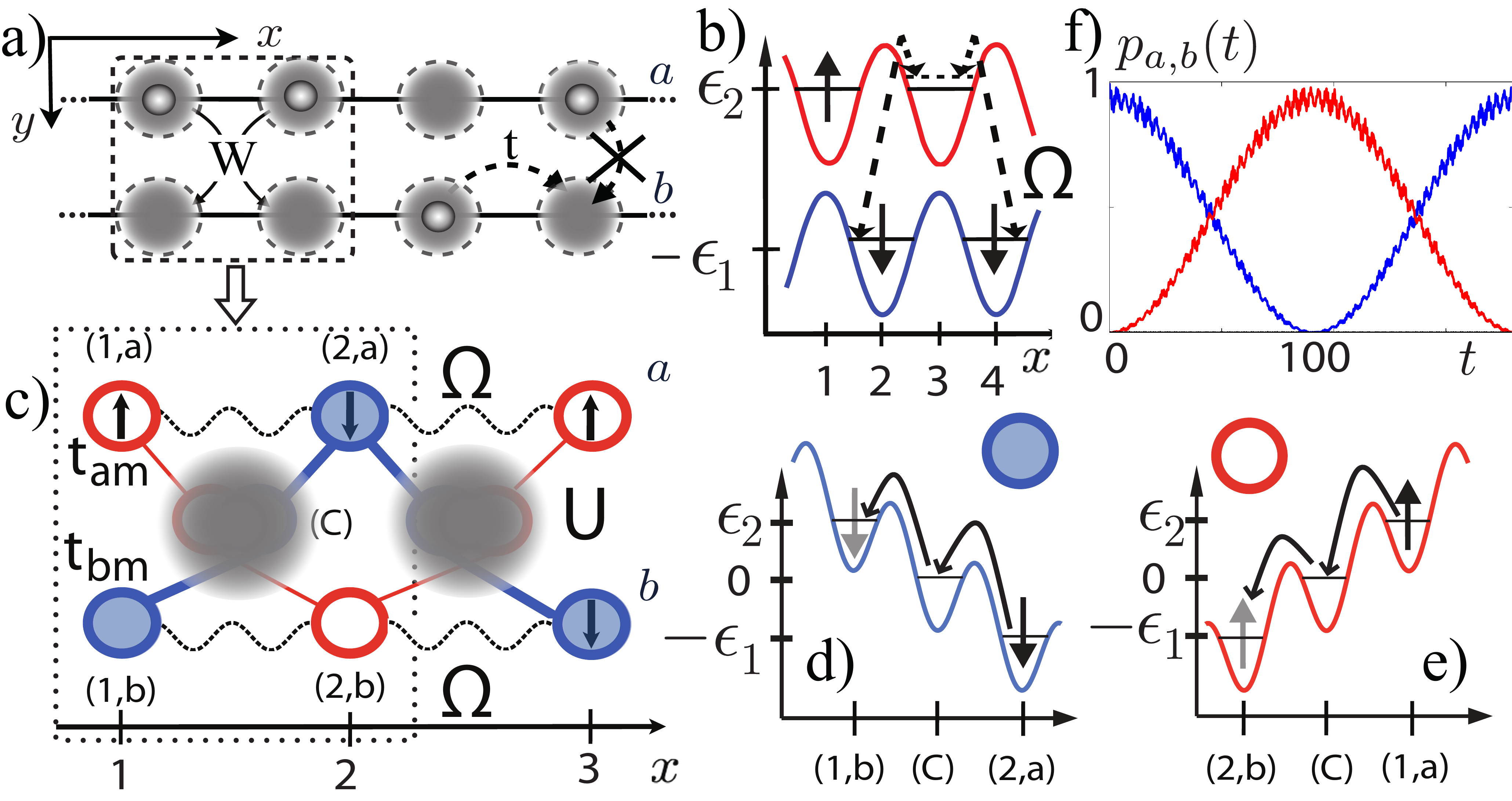}\caption{
\textit{a)} Ladder Hamiltonian: Atoms in the $a$- and $b$- wires can tunnel individually along the x-directions, and can hop in pairs between the wires. {\it b-e)} Implementation of the pair hopping:
{\it b)} The single wire is realized as a bipartite lattice of $\uparrow$ and $\downarrow$
fermions with Raman assisted tunneling (Rabi frequency $\Omega$).
{\it c)} Ladder scheme as a combination of two wires with opposite energy off-sets.
The dashed box denotes a single plaquette, with 
site indices indicated in parenthesis $(...)$; 
$t_{am}, t_{bm}$ are the tunneling amplitudes from the $a$- resp.
and $b$-wire to the central sites. Atoms in the center (C) of the plaquette
interact with strength $U$ (shaded areas). 
{\it d-e)} Energy off-sets along the diagonal 
of the plaquette in {\it c)} for the $\downarrow$ resp. $\uparrow$ species, 
and corresponding virtual processes indicating pair tunneling (see text).
{\it f)} Time evolution of the state $\ad_{1,a\uparrow}\ad_{2,a\downarrow}0\rangle$ according to the microscopic dynamics in units $t/t_{am}$
(see text): the blue (red) curve indicate the pair population $p_{a,b}(t)$
in the upper ($a$)/ lower ($b$) wire as a function of time.}%
\label{fig:cartoons}%
\end{figure}

In contrast, we propose and investigate in the present Letter an alternative
approach to create Majorana edge states in a purely number-conserving setting
\cite{cheng2011, fidkowski2011, sau2011}. We consider the conceptually
remarkably simple system of spinless fermions in two wires with
single-particle intrawire hopping, which are coupled via an \emph{interwire
pair hopping} (c.f. Fig.~\ref{fig:cartoons}a). On an intuitive level the
relation to Kitaev's model and existence of Majorana edge states is apparent,
when we consider one of the wires as an effective particle reservoir for the
second wire. The essential element in our system is pair hopping between the
wires, which breaks the $U(1)$ symmetry associated with the conservation of the particle-number difference between the
two wires, down to the $\mathbb{Z}_{2}$ parity symmetry, an ingredient known
to be crucial for the emergence of Majorana modes in the grand-canonical
scenario. The purpose of this work is two-fold. First, we provide evidence
for Majorana edge states and related topological order using both
field-theoretical arguments and detailed a Density-Matrix-Renormalization Group
(DMRG) study~\cite{White, Schollwoeck}. Second, we show that the present
setup with pair hopping has a natural implementation with cold atoms in
state-dependent optical lattices~\cite{bloch_review} combined with 
Raman assisted tunneling processes.

The emergence of Majorana edge states in the superfluid phase of the system is
demonstrated on the basis of the following criteria: (i) two degenerate ground
states with different parities for the individual wires in the case of open
boundary conditions (OBC), (ii) non-local fermionic correlations between the
edges, coming along with (iii) topological order indicated by a degenerate
entanglement spectrum, and (iv) robustness of the above properties against
static disorder. We also show that properties (i)-(iii) survive in the
presence of a weak single-particle hopping between the wires, also supporting the 
topological origin of the state. For experimental realizations, in particular
with atoms, the last property could be crucial, as it shows the robustness
against the most probable major imperfection.

\paragraph{Model.}

We consider the following Hamiltonian:
\begin{eqnarray}
H  &  =-\sum_{j}[(t_{a}a_{j}^{\dagger}a_{j+1}+t_{b}b_{j}^{\dagger}%
b_{j+1})+\text{h.c.}]\nonumber\label{Hlad}\\
&  +W\sum_{j}(a_{j}^{\dagger}a_{j+1}^{\dagger}b_{j}b_{j+1}+\text{h.c.}),
\end{eqnarray}
where $a_{j}$($a_{j}^{\dagger}$)$,$ $b_{j}$($b_{j}^{\dagger}$) are fermionic
annihilation (creation) operators defined on two distinct wires $a$ and $b$,
respectively, the first line describes intrawire single-particle hopping with
the corresponding amplitudes $t_{a,b}$ (in the following we consider
$t_{a}=t_{b}=t$ as a weak asymmetry of $t_{a,b}$ does not affect the results qualitatively), and the last term is the
\textsl{interwire} pair hopping with the amplitude $W$. The choice of the
Hamiltonian (1), motivated by previous considerations of the
number-conserving setting~\cite{cheng2011}, stems from global symmetries and
corresponding conserved quantities: Beside the total number of particles,
$N=N_{a}+N_{b}=\sum_{j}a_{j}^{\dagger}a_{j}+b_{j}^{\dagger}b_{j}$, associated
with the $U(1)$ symmetry, there is another conserved charge -- the
parity $P_{1}$ of one of the wires [say, the wire $a$, $P_{1}=p_{a}%
=(-1)^{N_{a}}$] associated with a $\mathbb{Z}_{2}$ symmetry~\cite{note1}.
The conservation is guaranteed by the last term in $H$ allowing only hopping
of particles between the wires in pairs, and is the key requirement to access
a topological phase with MFs.

Before presenting the analytical and numerical analysis of the Hamiltonian (1),
let us give an intuitive picture based on the simplest system supporting
fermionic Majorana edge states -- the 1D Kitaev quantum wire \cite{kitaev2001}
with $p$-wave pairing described by a mean-field BCS-like Hamiltonian resulting
from the coupling to a reservoir of Cooper pairs (see Ref.~\cite{kitaev2001}
for details). In our case, one could view one wire as a reservoir of pairs for
the other wire and vice versa, and decompose the pair-hopping term in a
mean-field manner as $W\sum_{i}(a_{i}^{\dagger}a_{i+1}^{\dagger}b_{i}%
b_{i+1}+\mathrm{h.c.})\rightarrow\sum_{i}(\Delta_{b}a_{i}^{\dagger}%
a_{i+1}^{\dagger}-\Delta_{a}^{\ast}b_{i}b_{i+1}+\mathrm{h.c.})$, where
$\Delta_{a}=W\left\langle a_{i}a_{i+1}\right\rangle $ and $\Delta
_{b}=W\left\langle b_{i}b_{i+1}\right\rangle $ are non-zero pairing amplitudes
which can be found by applying the standard Bogolyubov procedure. With this
decomposition, the Hamiltonian (1) describes two Kitaev wires
\cite{kitaev2001}, each of them having doubly-degenerate ground states with
different fermionic parities $p_{a,b}=\pm$\ for the $a$- and $b$-wire,
respectively, and carrying \textit{two} Majorana operators corresponding to
the edge-modes. Therefore, the ground state (GS) of the double-wire system (1)
with a fixed parity $P_{1}=p_{a}$ and a total parity $P=(-1)^{N}$ is doubly
degenerate. The two ground states can be connected by the product of
\textit{two} Majorana operators -- one from each wire. Strictly speaking,
long-wavelength fluctuations destroy long-range order in 1D breaking the
mean-field description even at zero temperature. In the considered case,
however, this does not change the picture qualitatively (see
Ref.\cite{fidkowski2011}).

\paragraph{Low-energy theory.}

Effective field theories based on bosonization~\cite{gogolin_book, giamarchi_book} represent a remarkable tool to
investigate the emergence of topological states and MFs in
strongly correlated systems \cite{stoudenmire2011,lobos2012,loss2011, Sela},
and has been applied recently to number conserving
settings~\cite{cheng2011,fidkowski2011}. Here, we employ this formalism to
qualitatively analyze the low-energy properties of the Hamiltonian
(\ref{Hlad}). We start with applying standard bosonization formulas to
introduce effective low-energy phase and density fluctuation fields
$\varphi_{\gamma},\vartheta_{\gamma}$, respectively, for each species
$\gamma=a,b$~\cite{supmat}. After introducing symmetric and antisymmetric
combinations, $\varphi_{S/A}=(\varphi_{a}\pm\varphi_{b})/\sqrt{2}$, and
neglecting contributions with high scaling dimensions, the bosonized
Hamiltonian decouples into symmetric and antisymmetric sectors. The symmetric
sector describes collective density-wave excitations, and is well-captured by
a Tomonaga-Luttinger liquid Hamiltonian:
\begin{equation}
H_{S}=\frac{v_{S}}{2}\int\left[  \frac{(\partial_{x}\varphi_{S})^{2}}{K_{S}%
}+K_{S}(\partial_{x}\vartheta_{S})^{2}\right]  dx,
\end{equation}
whilst the antisymmetric one is described by a sine-Gordon Hamiltonian
\cite{gogolin_book}:
\begin{equation}
H_{A}=\frac{v_{A}}{2}\int\left[  \frac{(\partial_{x}\varphi_{A})^{2}}{K_{A}%
}+K_{A}(\partial_{x}\vartheta_{A})^{2}+w\cos[\sqrt{4\pi}\vartheta_{A}]\right]
dx, \label{Hanti}%
\end{equation}
where $K_{\alpha}$ and $v_{\alpha}$ are the Luttinger parameter and the sound
velocity, respectively, for each sector $\alpha=(A,S)$, and $w\propto W$
results from the pair hopping. It can be shown that the parity symmetry
$\mathbb{Z}_{2}$ and the number conservation are exactly retained at low
energies in the antisymmetric and symmetric sector,
respectively~\cite{gogolin_book,cheng2011}. We now discuss the qualitative
phase diagram of the system by using standard Renormalization Group (RG)
scaling arguments~\cite{gogolin_book, giamarchi_book, difrancesco}. Away from
the strong coupling limit $W\gg t$ (where terms with higher scaling dimensions
may become relevant), the two sectors remain decoupled, so that one can
analyze them separately. While the symmetric sector is simply a theory of free
bosons, the antisymmetric sector displays richer physics, as it undergoes a
phase transition from a gapless phase at $W=0$ to a gapped, superconducting
phase for $W>0$. In analogy with the continuum model of Ref.~\cite{cheng2011},
Eq.~(\ref{Hanti}) can be exactly mapped to the continuum version of the Kitaev
wire \cite{cheng2011} at the Luther-Emery point $K_{A}=2$. As a result, the
system with OBC displays a two-fold ground state degeneracy, where the two
states have opposite parities $P_{1}$, and support MFs at the
boundaries~\cite{supmat}. Moreover, the single-particle correlation functions show
exponential decay $\langle a_{i}^{\dagger}a_{i+x}\rangle\simeq e^{-\xi
\left\vert x\right\vert }$ in the bulk, signaling the presence of a finite
superconducting gap. 
Away from the Luther-Emery point, the MF wave function overlap
increases depending on $(K_A-2)$, the corresponding splitting in the 
GS degeneracy at finite system size being $e^{-\kappa L}$, 
eventually turning to power-law in the presence of certain kinds of perturbations
\cite{cheng2011,fidkowski2011,sau2011}.
On the other hand, in
the strong coupling limit $|W|\gg t$, the presence of additional terms (with
higher scaling dimension) of the form $w\cos[\sqrt{4\pi}\vartheta
_{A}](\partial_{x}\varphi_{S})^{2}\simeq-w(\partial_{x}\varphi_{S})^{2}$ leads
to a reduction of the sound velocity $v_{S}$, resulting in
phase separation.

\paragraph{Numerical results}

Employing this low-energy picture as a guide, we now present a quantitative
numerical investigation of the Hamiltonian Eq.~(\ref{Hlad}). We start with a
brief description of the phase diagram of the system, and then discuss the
criteria (i)-(iv) relevant for the existence of MFs. In the
following, we set $t=1$ as the energy scale.

The phase diagram of the model can be divided into three regions: a
superconducting phase, an insulating phase, and a region of phase separation.
The superconducting phase is characterized by a homogeneous density, leading
superconducting correlations, and nonzero single-particle gap $\Delta
=|E_{0}(N)-\tfrac{1}{2}(E_{0}(N+1)+E_{0}(N-1))|$ for periodic boundary
conditions (PBC). Here $E_{0}(N)$ is the ground state energy for $N$
particles. We find this phase for small and moderate values of the pair
hopping $|W|\gtrsim1$ and all fillings except $n=1/2$. At exactly
half-filling, an incompressible insulating phase is formed with exponentially
decaying superconducting correlations. For large values of the pair hopping
$|W|\gg1$ we find phase separation with the formation of particle clusters. In
the following we concentrate on the superconducting phase and check the
criteria (i)-(iv). For our numerical analysis, we take $W=-1.8$ and the
filling $n=1/3$ as representative values resulting in a homogeneous
superconducting phase for system sizes $L=12$, $24$ and $L=36$ with even
number of particles.

\begin{figure}[t]
\includegraphics[width = 0.98\columnwidth]{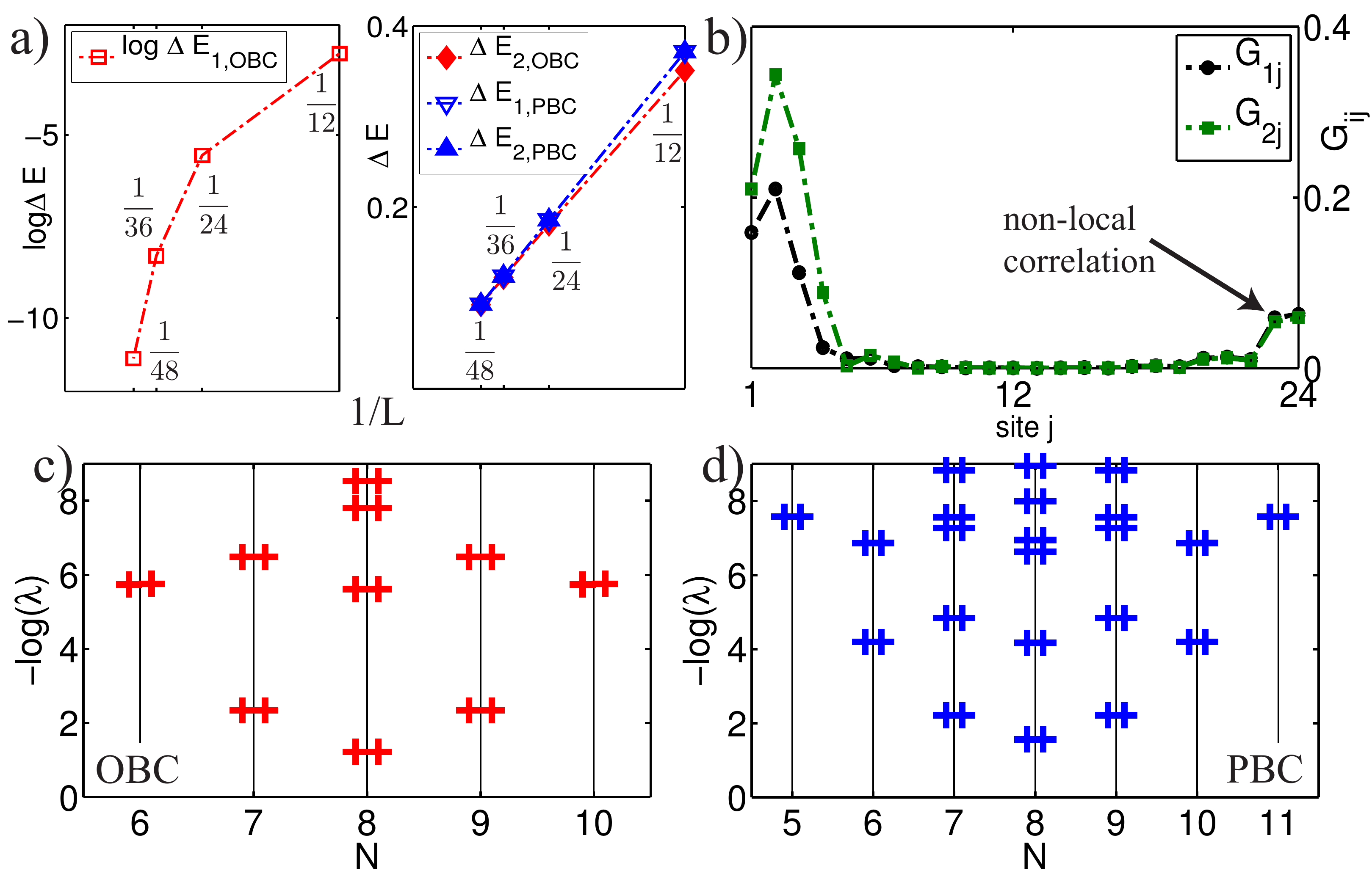}\caption{a)
Closing of the energy gaps with the system size ($L=12,24,36$) for $W=-1.8$
and $n=1/3$. For OBC, the gap $\Delta E_{1}$ closes exponentially (left
panel), in contrast to the polynomial closing in the case of PBC (right panel,
open blue triangles). The energy gap $\Delta E_{2}$ closes polynomially
independent of the boundary conditions (right panel, red diamonds for OBC and
closed triangles for PBC). Note that $\Delta E_{2,\mathrm{PBC}}=\Delta
E_{1,\mathrm{PBC}}$. b) Non-local fermionic correlations $G_{lj}$ on the upper
wire for $L=24$. c) and d) Entanglement spectrum for the system of the size
$L=24$ shows double degeneracy for both OBC (c) and PBC (d).}%
\label{fig:LadderPerfect}%
\end{figure}

(i) The ground state degeneracy can be studied by looking at the energy gap
$\Delta E_{n}(N)=E_{n}(N)-E_{0}(N)$ between the ground state and the $n$-th
excited state. As shown in Fig.~\ref{fig:LadderPerfect}a, in the case of OBC,
the gap between the ground and the first excited state $\Delta
E_{1,\mathrm{OBC}}$ closes exponentially in the system size (left panel)
indicating the degeneracy of the ground state in the thermodynamic limit. This
is in contrast to the case of PBC that is depicted in the right panel of
Fig.~\ref{fig:LadderPerfect}a (blue open triangles). Here we find that $\Delta
E_{1,\mathrm{PBC}}$ closes linearly in the system size, and $\Delta
E_{1,\mathrm{PBC}}=\Delta E_{2,\mathrm{PBC}}$, i.e. the first and second
excited state are degenerate (blue open and closed triangles). For OBC, $\Delta
E_{2,\mathrm{OBC}}$ also closes linearly in the system size (red diamonds). We
find that the two degenerate ground states in the case of OBC differ by the
parities of the individual wires. Note that
for OBC we also have $\Delta=0$.

(ii) The intrawire single-particle correlation function $G_{lj}=\langle
a_{l}^{\dagger}a_{j}\rangle$ for the system of the length $L=24$ is shown in
Fig.~\ref{fig:LadderPerfect}b for the case where $l=1,2$ is close to the left
edge and $j\in\lbrack l,L]$. We see that $G_{lj}$, being exponentially small
inside the wire, attains a finite value at the right edge showing the
existence of non-local fermionic correlations typical for a system with MF
edge states.

(iii) Topological order (TO) manifests itself in the degeneracy of the
entanglement spectrum (ES)~\cite{Pollmann,Turner,Fidkowski_ES}: Let $\rho
_{A}=\sum_{Nj}\lambda_{j}^{(N)}\rho_{j}^{(N)}$ be the reduced density matrix
of the system with respect to some bipartition with support on both wires,
where $\rho_{j}^{(N)}$ describes a pure state of $N$ particles with the
corresponding eigenvalues $\lambda_{j}^{(N)}$. In a topological phase, the
low-lying eigenvalues $\lambda_{j}^{(N)}$ are expected to be doubly degenerate
for each $N$, for both OBC and PBC, as it is
demonstrated in Figs.~\ref{fig:LadderPerfect}c (OBC) and
\ref{fig:LadderPerfect}d (PBC) for a system of the size $L=24$. Moreover, the
distributions of the low-lying eigenvalues as a function of $N$ share the same
patter in the two cases. 

\begin{figure}[t]
\includegraphics[width = 0.95\columnwidth]{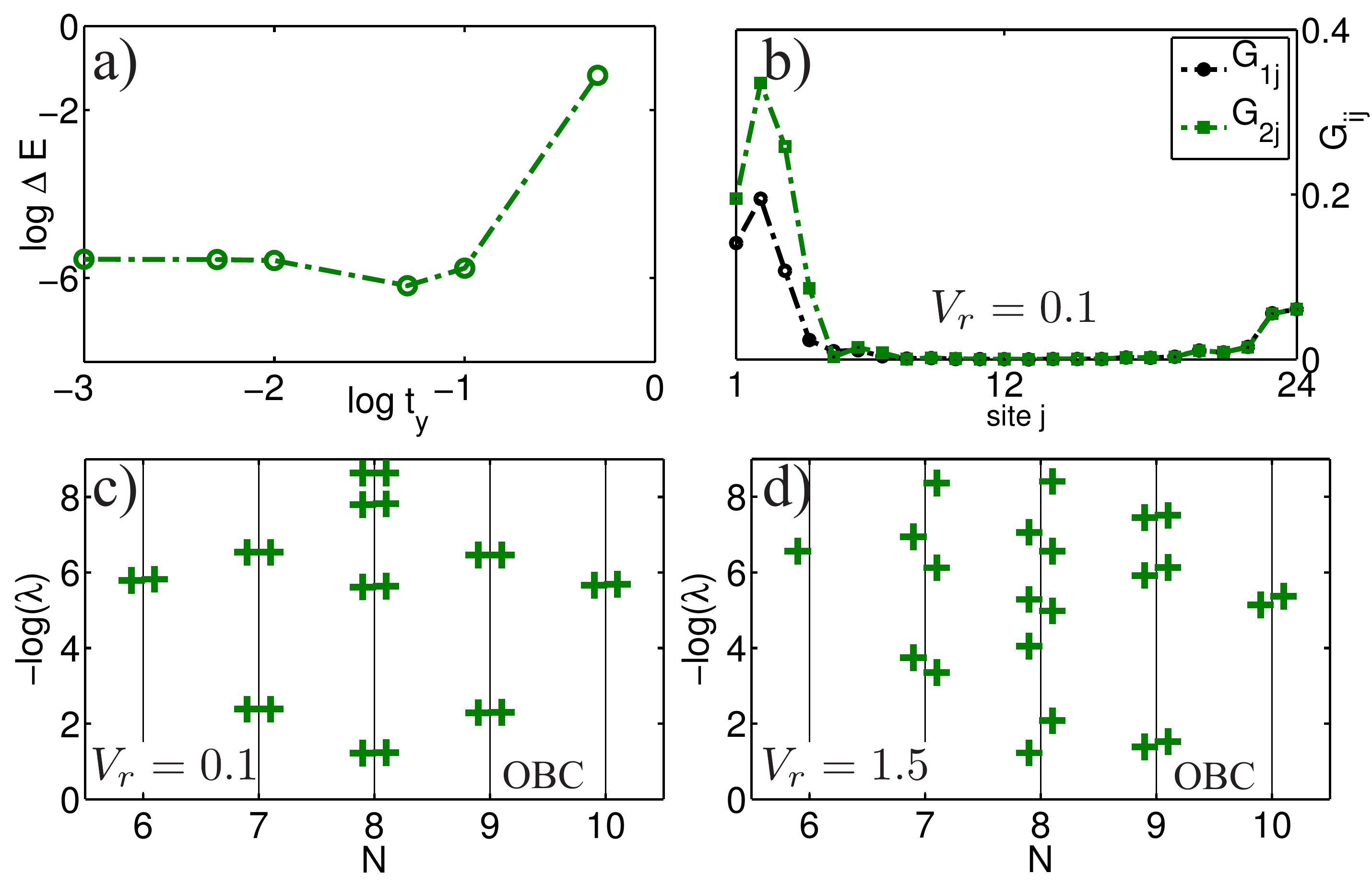}\caption{Effects
of imperfections on the topological order ($L=24,$ $n=1/3$).
a) Ground state degeneracy in the presence of an interwire
single-particle hopping $H_{\perp}=\sum_{i}t_{y}a_{i}^{\dagger}b_{i}+h.c.$. 
b)-d) Effects of
static disorder: The non-local correlations (b) and the degeneracy of the ES
(c) indicated the topological state in the presence of disorder with
$V_{r}=0.1t$. d) Breaking of the topological phase by a strong disorder with
$V_{r}=1.5t$.}%
\label{fig:LadderNonperfect}%
\end{figure}

(iv) The robustness of the above properties against static disorder is one of
the key manifestations of a non-local topological order. We model the disorder
by adding the term $H_{V_{r}}=\sum_{j}V_{j}^{(a)}a_{j}^{\dagger}a_{j}%
+V_{j}^{(b)}b_{j}^{\dagger}b_{j}$ to the Hamiltonian, where $V_{j}^{(\gamma)}%
$with $\gamma=a,b$ are random local potentials equally distributed in the
interval $[-V_{r},V_{r}]$. We find that even for moderate disorder
$V_{r}=0.1t$, the ground state remains doubly degenerate, and the system still
exhibits the non-local correlations (Fig.~\ref{fig:LadderNonperfect}b) as well
as the degenerate ES (Fig.~\ref{fig:LadderNonperfect}a), indicating the
presence of topological order. For strong local disorder, however, the
topological effects disappear, as exemplified by the non-degenerate ES for
$V_{r}=1.5t$ in Fig.~\ref{fig:LadderNonperfect}d.

Remarkably, the observed topological order and its consequences are also
stable against a single-particle hopping $H_{\perp}=\sum_{i}t_{y}%
a_{i}^{\dagger}b_{i}+h.c.$ between the two wires, which breaks the
parity\textbf{ }of the wires and related $Z_{2}$ symmetry. As an example, in
Fig.~\ref{fig:LadderNonperfect}a we show the energy gap $\Delta
_{1,\mathrm{OBC}}$ as a function of $t_{y}$: The ground state of the system
remains quasi-degenerate ($\Delta E\simeq 10^{-5}$) up to values $t_{y}$ of the order of $0.1t$, in agreement
with the prediction of Refs. \cite{cheng2011, fidkowski2011}. Note, however,
that the dependence of $\Delta_{1,\mathrm{OBC}}$ on $L$ changes from
exponential to power law \cite{sau2011}. This stability could be very
important for experimental realizations of the model because the interwire
single-particle hopping is one of the most probable imperfections.

\paragraph{Pair hopping with cold fermionic atoms.}
 The key ingredient of the Hamiltonian
(\ref{Hlad}) is the interwire pair hopping with coupling $W$ in the absence of (parity
violating) single particle tunneling. The basic idea behind an atomic
implementation is to introduce offsets in optical lattices, which suppress
single particle hopping by energy constraints, while an energy conserving pair
hopping is allowed and mediated by interactions. 

An atomic setup illustrating these ideas is given in Fig.~\ref{fig:cartoons}, while technical
details and variants of the scheme can be found in the SI.
We implement the two wires of spinless fermions as a bipartite lattice for
spinful fermions. Odd and even lattice sites $j$ trap
the spin $\uparrow$ and $\downarrow$ components of the fermions with energies
$\epsilon_{2}$ and $-\epsilon_{1}$,  respectively, and
transitions between the adjacent wells are induced by an external RF field or
a Raman assisted hopping (c.f Fig.~\ref{fig:cartoons}b). This realizes the first line of $H$ in
Eq.~(\ref{Hlad}). To understand the pair hopping mechanism, consider the plaquette
indicated in Fig.~\ref{fig:cartoons}c by the dashed line. We assume an auxiliary molecular site
in the center of the plaquette (indicated as $(\textrm{C})$ in Fig.~\ref{fig:cartoons}c), which traps both $\uparrow$ and
$\downarrow$ atoms, and is connected to the lattice sites on the wire by a
spin-preserving tunneling coupling with amplitudes $t_{{am}}$ and $t_{bm}$. Pairs of atoms occupying the molecular
site are assumed to interact via an onsite interaction $U$. In addition, we
introduce spin-dependent lattice offsets, which are indicated by the
$-\epsilon_1, \epsilon_2$ for the lattice sites on the two wires and 
for $\uparrow$ and $\downarrow$ species, respectively (Fig.~\ref{fig:cartoons}d-e). Such offsets can be generated as Zeeman shifts of the spin
states, if a gradient magnetic field is applied perpendicular to the wire. 

Single particle hopping between the wires is suppressed in this setup:
consider an atom, say in the upper wire $a$ in lattice site $1$ with spin
$\uparrow$. Spin-preserving tunneling is possible via the molecular site along the
diagonal of the plaquette (virtual processes are indicated in Fig.~\ref{fig:cartoons}d-e).
It  corresponds to the process
$\uparrow_{1a}\rightarrow\uparrow_{m}\rightarrow\uparrow_{2b}$, which is
suppressed by the corresponding energy offsets $+\epsilon_2,0,-\epsilon_1$. In a
similar way also the tunneling of the $\downarrow$ atom along $\downarrow
_{2a}\rightarrow\downarrow_{m}\rightarrow\downarrow_{1b}$ is suppressed by
energy conservation. However, for pair hopping $\uparrow
_{1a}\downarrow_{2a}\rightarrow\uparrow\downarrow_{m}\rightarrow\uparrow
_{2b}\downarrow_{1b}$ the overall energy will be conserved, since the two atoms can
exchange energy via the interaction $U$. After adiabatic elimination of the 
intermediate sites when $U, \epsilon_{1/2}\gg t_{am, bm}$, the resulting amplitude  for the pair-hopping term is 
$W\simeq t_{am}^{2}%
t_{bm}^{2}(1/\epsilon_1-1/\epsilon_2)^2/U$  (see SI). Note that the pair-hopping
process $\uparrow_{1a}\downarrow_{1b}\rightarrow\downarrow_{2a}\uparrow_{2b}$
will also be allowed, but does not change the number of particles on the
wires, and thus preserves atom number parity on the wires. A detailed description of this pair hopping dynamics including possible imperfections, e.g. induced by the Raman couplings, can be found in the SI. In Fig.~\ref{fig:cartoons}f we present a numerical analysis of the pair hopping dynamics, where (in units of $t_{am}=t_{bm}=1$) $\epsilon_2 = 2\epsilon_1 = 2$,  $U = -20$ (see SI).  Finally, note that the engineering of pair hopping has further applications in cold atom systems. For example, the pair hopping can be used as an entangling quantum gate, where the hopping of one particle
(control) triggers the tunneling of a second atom (target). Further, it has applications in the context of quantum simulation, e.g. for lattice gauge theories emulation include ring-exchange
and {\it rishon} determinant interactions~\cite{buechler2005,banerjee2012}.

\paragraph{Detection.}

Finally, we address the problem of detecting the emerging Majorana states in
our AMO setup. Following the proposals of Ref. \cite{kraus2012}, this could be
done, e.g., by using standard AMO detection tools like time-of-flight imaging
and spectroscopic techniques to probe the ground state degeneracy and the
inherent non-local fermionic correlations. Demonstration of a non-Abelian
statistic of the MFs, on the other hand, requires some dynamical protocols
resulting in the motion of MFs around each other. In our setup, one could
think of a generalization of the ideas of Ref. \cite{kraus2012b} relying on
single-site addressing available in current experiments with ultra-cold
atoms~\cite{greiner2009,kuhr2010}. Another possibility would be an atomic
analog of the fractional Josephson effect~\cite{Rokhinson} using a properly
shaped external potential along the $x$-direction.

\paragraph{Conclusions.}

In summary, we have shown that topological states of matter with Majorana fermion
edge states can be created in fermionic atomic ladders without any additional
reservoir or p-wave interaction, but with only interwire pair hopping, which
could provide an easier, complementary way for experimental realizations.

\paragraph{Acknowledgments.}

We thank N. Ali-Bray, M. Burrello, S. Manmana, J. D. Sau, F. Schreck and H.-H.
Tu for fruitful and useful discussions. M.D. acknowledges support by the
European Commission via the integrated project AQUTE. We further acknowledge
support by the Austrian Science Fund FWF (SFB FOQUS F4015-N16) and the
Austrian Ministry of Science BMWF as part of the UniInfrastrukturprogramm of
the Research Platform Scientific Computing at the University of Innsbruck.

\cleardoublepage

\section{SI: Pair hopping model: low-energy field theory}

We present here some details of the low-energy field theory in the lattice
pair-hopping model presented in the main text. The Hamiltonian is (for
compactness, we adopt here a different notation $c_{a,j}=a_{j}$,
$c_{b,j}=b_{j}$ with respect to the main text)
\begin{align}
H &  =-\sum_{j;\gamma=a,b}t_{\gamma}(c_{\gamma,j}^{\dagger}c_{\gamma
,j+1}+h.c.)+\nonumber\label{eq_H}\\
&  +W\sum_{j}(c_{a,j}^{\dagger}c_{a,j+1}^{\dagger}c_{b,j}c_{b,j+1}+h.c.)
\end{align}
and has a $U(1)\otimes\mathbb{Z}_{2}$ symmetry conserving both the total
number of particles and the parity of each wire. We apply here the standard
bosonization lattice procedure in order to extract the low-energy field theory
\cite{gogolin_bookbib, giamarchi_bookbib}. Within this framework, we take first the
continuum limit ($a_{0}$ being the lattice spacing) and introduce
\begin{equation}
c_{j,\gamma}^{\dagger}=\psi_{R,\gamma}^{\dagger}(x=ja_{0})+\psi_{L,\gamma
}^{\dagger}(x=ja_{0}).
\end{equation}
The bosonization identities have the following form:
\begin{align}
\psi_{r,\gamma}(x) &  =\frac{\eta_{r,\gamma}}{\sqrt{2\pi a_{0}}}%
e^{irk_{F,\gamma}x}e^{-i[r\varphi_{\gamma}-\vartheta_{\gamma}]}=\nonumber\\
&  =\frac{\eta_{r,\gamma}}{\sqrt{2\pi a_{0}}}e^{irk_{F,\gamma}x}e^{-\frac
{i}{\sqrt{2}}[r\varphi_{S}-\vartheta_{S}+\bar{\gamma}(r\varphi_{A}%
-\vartheta_{A})]},
\end{align}
where $R/L$ implies $r=\pm1$ and $\bar{\gamma}(a/b)=\pm1$, and we have
defined
\[
\vartheta_{A}=\frac{\vartheta_{a}-\vartheta_{b}}{\sqrt{2}},\quad\vartheta
_{S}=\frac{\vartheta_{a}+\vartheta_{b}}{\sqrt{2}},
\]
and the Klein operators
\begin{equation}
\{\eta_{R,\gamma},\eta_{L,\gamma}\}=0,\quad\lbrack\eta_{r,a},\eta_{s,b}]=0,
\end{equation}%
\begin{equation}
\{\eta_{r,\gamma}^{\dagger},\eta_{s,\beta}\}=2\delta_{sr}\delta_{\gamma\beta
},\quad\eta_{s,\gamma}\eta_{s,\gamma}^{\dagger}=1
\end{equation}
in order to preserve fermionic commutation relations. The free part (the first
line) of the lattice Hamiltonian is then $H_{A}+H_{S}$, where ($\Delta=A/S$)
\begin{equation}
H_{\Delta}=\frac{v_{\Delta}}{2}\int\left[  \frac{(\partial_{x}\varphi_{\Delta
})^{2}}{K_{\Delta}}+K_{\Delta}(\partial_{x}\vartheta_{\Delta})^{2}\right]  ,
\end{equation}
with Tomonaga-Luttinger parameters $K_{\Delta}=1$ and sound velocities
$v_{\Delta}=2ta_{0}^{-2}\sin(k_{F,\Delta}a_{0})$ with $k_{F,S}=\pi n$. The
bosonized pair-hopping operator [the second line in Eq.~\eqref{eq_H}] reads
\begin{align}
H_{W}^{(A)} &  =\frac{4W}{(2\pi a_{0})^{2}}\int\frac{dx}{a_{0}}\Big[\gamma
(n)\cos[\sqrt{8\pi}\vartheta_{A}(x)+\nonumber\\
&  +a_{0}^{-2}\cos[\sqrt{8\pi}\vartheta_{A}(x)]\cos[a_{0}k_{S}](\partial
_{x}\varphi_{S})^{2}\Big],
\end{align}
where $\gamma(n)$ is a density-dependent coefficient and we omit the Klein
factors. Away from the strong coupling limit $W/t\gg1$, where interactions
between symmetric and (gapped) antisymmetric sector drive the system towards
phase separation, the bosonized Hamiltonian $H$ can be effectively split into
two distinct sectors. The symmetric one is described by a Tomonaga-Luttinger
liquid, whereas the antisymmetric one by a sine-Gordon Hamiltonian with \ the
mass-term for the phase field $\vartheta_{A}$. In particular, at the
Luther-Emery point $K=2$, it is possible to map the antisymmetric sector to
the so called massive Dirac Hamiltonian \cite{cheng2011bib}
\begin{equation}
\tilde{H}=-iv_{A}(\xi_{R}^{\dagger}\partial_{x}\xi_{R}-\xi_{L}^{\dagger
}\partial_{x}\xi_{L})+im(\xi_{R}^{\dagger}\xi_{L}^{\dagger}-\xi_{L}\xi
_{R}),\label{H_dirac}%
\end{equation}
by employing standard re-fermionization techniques
\cite{gogolin_bookbib,vondelft1998}. Here $\xi_{s}(s=L/R)$ are Dirac fermions
with mass $m\sim W$, which are related to the bosonic operators via $\xi
_{s}=\eta_{R,b}^{\dagger}\eta_{L,a}e^{i\sqrt{4\pi}r_{s}\varphi_{s,A}}$, where
$r_{s=R/L}=\pm1$ and the chiral fields $\varphi_{s=R/L,A}=(\varphi_{A}%
/\sqrt{2}\pm\sqrt{2}\vartheta_{A})$. The mapping to Eq.~\eqref{H_dirac} is
possible due to the specific form of the interaction in the sine-Gordon
Hamiltonian involving phase fluctuations $\vartheta_{A}$. The effective
Hamiltonian \eqref{H_dirac} corresponds to the continuum limit of the Kitaev
chain~\cite{cheng2011bib}, and, therefore, supports Majorana edge states. Since
the antisymmetric and symmetric sectors are decoupled, the energy splitting
between the two degenerate ground states is exponentially small in the system
size \cite{fidkowski2011bib, cheng2011bib}. However, certain external perturbations,
such as interchain backscattering (in the presence of interactions between the
wires) or impurities, may change the dependence of the energy splitting on the
system size from exponential to algebraic (although with an extremely small
prefactor in the vicinity of the Luther-Emery point, see the detailed
instanton discussion in Ref.~(\onlinecite{fidkowski2011bib}).

\section{SI: Effective pair-hopping Hamiltonian - Model 1}

In this and the following Section, we describe two possible implementation
schemes realizing the pair hopping Hamiltonian discussed in the main text.

\subsection{Single Wire}

Before discussing the two couple wires and pair hopping between these wires we
briefly describe our setup for a single wire.

We implement the single wire of spinless fermions as a bipartite lattice of
spinful fermions in an 1D-optical lattice as indicated in
Fig.~\ref{fig:model2}a. Odd and even lattice sites $j$ trap $\uparrow$ and
$\downarrow$ components of fermionic atoms, respectively, and transitions
between adjacent wells (and the associated spin flip) are induced either by an
RF field or by an optical Raman transition. \ The corresponding Hamiltonian is%
\begin{align*}
H  &  =\epsilon_{2}\sum_{j\text{ odd}}a_{j\uparrow}^{\dagger
}a_{j\uparrow}-\epsilon_{1}\sum_{j\text{ even}}a_{j\downarrow
}^{\dagger}a_{j\downarrow}\\
&  -\Omega\sum_{r=\pm1;j\text{ even}}\left(  a_{j+r\downarrow}^{\dagger
}a_{j\downarrow}e^{-i\omega t}+\mathrm{h.c.}\right).
\end{align*}
with $\epsilon_{\sigma=1,2}$ the energies of the atomic states
representing the spins, $\Omega$ the Rabi frequency of the Raman drive and
$\omega$ the corresponding frequency. After a transformation to a \ "rotating
frame" the above Hamiltonian can be rewritten with the mapping $a_{j\downarrow
}\rightarrow a_{j}$ ($j$ even) and $a_{j\uparrow}\rightarrow a_{j}$ ($j$ odd)
as%
\[
\tilde{H}=-t\sum_{r=\pm1;j}a_{j+r}^{\dagger}a_{j}+(-\delta)\sum_{j\text{ odd}%
}a_{j}^{\dagger}a_{j}%
\]
Here $\delta=\omega-(\epsilon_{1}+\epsilon_{2})$ is the
detuning , which acts as an offset (superlattice) for the odd lattice sites.
The tunneling amplitude in the lattice can be identified with the Rabi
frequency of the drive, $t\equiv\Omega$. Note that the above Hamiltonian
corresponds to an effective model of spinless fermions hopping on a 1D lattice.

\subsection{Two Coupled Wires}

\begin{figure}[t]
\includegraphics[width = 0.95\columnwidth]{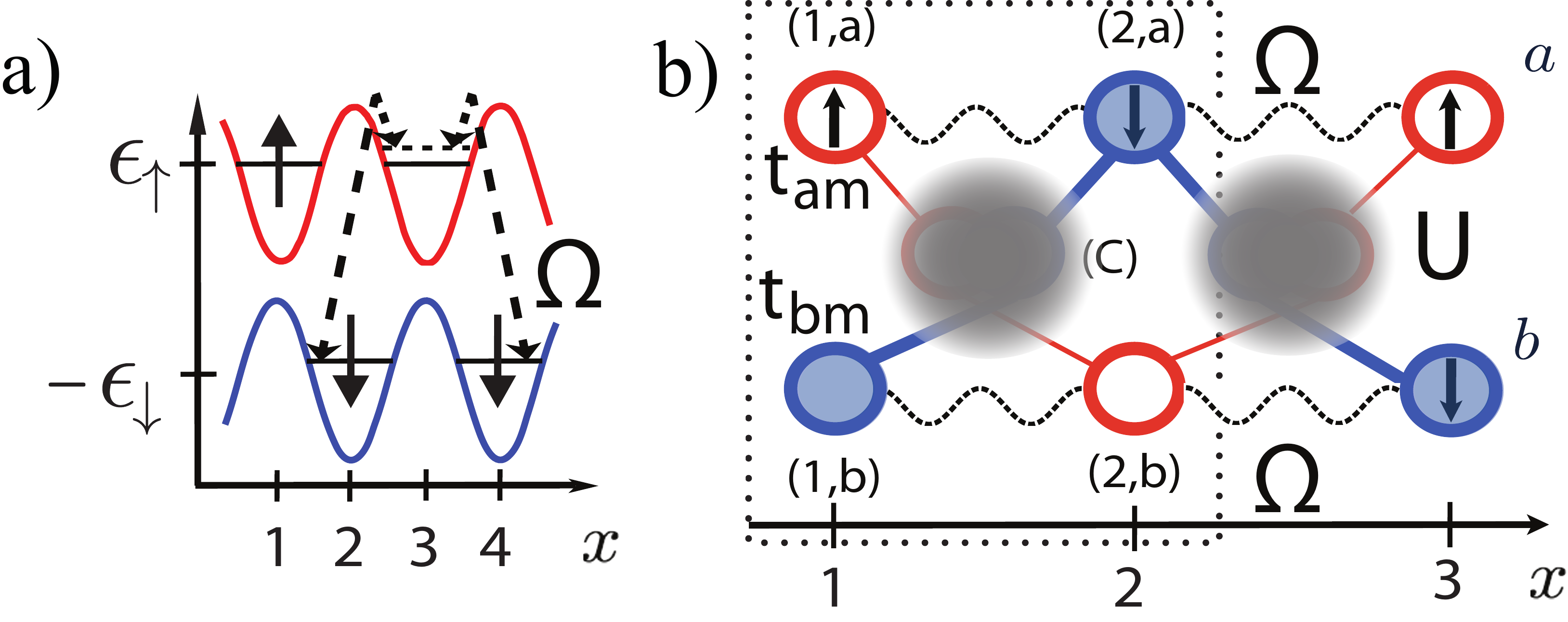} \newline%
\caption{Microscopic illustration of Model I. Panel \textit{a)}: the single
($a$-)wire described by $\tilde{H}$ is obtained by combining a species
dependent lattice for the $\uparrow$ and $\downarrow$ species, with Raman
assisted tunneling between the two, the small detuning thereof providing a
finite value of the $\delta$ parameter. Panel \textit{b)}: two-wire setup of
the Hamiltonian $\tilde{H}_{\square}$, indicating the correspondent energy
off-set of each site. The upper and lower wires are denoted as $a$ and $b$,
respectively. Here, in the single plaquette included in the dotted box, thin
lines link sites connected by standard tunneling, and dashed curved lines link
sites connected by Raman assisted tunneling. }%
\label{fig:model2}%
\end{figure}

The above realization of a 1D wire of spinless fermionic atoms acts as a
building block for two coupled wires (with a straightforward generalization to
many coupled wires). The setup we have in mind is depicted in
Fig.~\ref{fig:model2}b with the wires denoted by $a$ and $b$, respectively. In
addition we introduce auxiliary sites between the wires, which we call center
or molecular sites in Fig.~\ref{fig:model2}b. They are the basic ingredient in
implementing pair hopping, while strongly suppressing unwanted processes like
single particle tunneling (which violates parity of atoms on the two wires).

We extend the single wire model of the previous subsection assuming that the
spins are placed in a spatially varying magnetic field with gradient
perpendicular to the wires. This results in a spin dependent energy offset:
the spins $\uparrow$ and $\downarrow$ on the (upper) $a$ wire
have Zeeman energies $\epsilon_2$ and $-\epsilon_1$,
and the corresponding energies on the (lower) wire $b$ are
$-\epsilon_1$ and $\epsilon_2$,
respectively. Thus we have the following Hamiltonians for the two (uncoupled) wires:%

\begin{align*}
H_{0a}  &  =\epsilon_{2}\sum_{j\text{ odd}}a_{j\uparrow
,a}^{\dagger}a_{j\uparrow,a}-\epsilon_{1}\sum_{j\text{
even}}a_{j\downarrow,a}^{\dagger}a_{j\downarrow,a}\\
&  -\sum_{r=\pm1;j\text{ even}}\left(  \Omega a_{j+r\uparrow,a}^{\dagger
}a_{j\downarrow,a}e^{-i\omega t}+\mathrm{h.c.}\right)  ,\\
H_{0b}  &  =\epsilon_{2}\sum_{j\text{ odd}}a_{j\downarrow
,b}^{\dagger}a_{j\downarrow,b}-\epsilon_{1}\sum_{j\text{
even}}a_{j\uparrow,b}^{\dagger}a_{j\uparrow,b}\\
&  -\sum_{r=\pm1;j\text{ even}}\left(  \Omega a_{j+r\downarrow,b}^{\dagger
}a_{j\uparrow,b}e^{+i\omega t}+\mathrm{h.c.}\right)  .
\end{align*}
If we choose $\omega=\epsilon_{1}+\epsilon_{2}$, our
model reduces again to a 1D tight binding model for spinless fermions hopping
on the wires, as discussed in the previous subsection. The Zeeman offsets will
play a central role in the following discussion of pair hopping.

As shown in Fig.~\ref{fig:model2}b, atoms can hop from the two wires to
(auxiliary) central sites, which can be occupied by both $\uparrow$ and
$\downarrow$ atoms. Two atoms occupying a central site will interact according
to an onsite interaction $U$, effectively forming a "molecule" $m$. The
corresponding Hamiltonian is
\[
H=H_{0}+H_{1}=H_{0a}+H_{0b}+H_{0c}+H_{1}.
\]

Here $H_{0a}$ and $H_{0b}$ are Hamiltonians for the wire as discussed above.
The Hamiltonian for the central sites $c$ is
\begin{align*}
H_{0c}=  &  \sum_{c\equiv(j,j+1)}Ua_{\uparrow c}^{\dagger
}a_{\downarrow c}^{\dagger}a_{\downarrow c}a_{\uparrow c}\\
&  -\sum_{c\equiv(j,j+1)}\sum_{\alpha=a,b}t_{m\alpha}\left(  a_{\uparrow
c}^{\dagger}a_{j\uparrow,\alpha}+a_{\downarrow c}^{\dagger}a_{j+1\downarrow
,\alpha}+\mathrm{h.c.}\right)
\end{align*}
where we adopt the notation $c\equiv(j,j+1)$ for $c$ on the link $j,j+1$,
and we have an interaction term $U$ between atoms with different spin
in the center of the plaquette. The last line is a spin-preserving hopping from the
wires to the central site.

Finally, $H_{1}$ accounts for RF or Raman induced spin-flip transitions,
\begin{align*}
H_{1} &  =-\sum_{c\equiv(j,j+1)}\left(  \Omega^{\prime}e^{-i\omega
t}a_{\uparrow c}^{\dagger}a_{\downarrow c}+\mathrm{h.c.}\right)  \\
&  -\sum_{j\text{ even}}\Omega^{^{\prime\prime}}\left(  a_{j\uparrow,a}^{\dag
}a_{\uparrow c}+a_{j\downarrow,b}^{\dagger}a_{\downarrow c}+\right.\\
&+\left.a_{\downarrow
c}^{\dag}a_{j+1\uparrow,a}+a_{\uparrow c}^{\dag}a_{j+1\uparrow,b}\right)
e^{-i\omega t}+\mathrm{h.c.}\\
&  -\sum_{j\text{ odd}}\Omega^{^{\prime\prime}}\left(  a_{j\uparrow,a}^{\dag
}a_{\uparrow c}+a_{j\downarrow,b}^{\dagger}a_{\downarrow c}+\right.\\
&+\left.a_{\downarrow
c}^{\dag}a_{j+1\uparrow,a}+a_{\uparrow c}^{\dag}a_{j+1\uparrow,b}\right)
e^{+i\omega t}+\mathrm{h.c.~.}%
\end{align*}
The first line is a spin-flip at the central site, and the last
lines account for an RF or Raman assisted hopping from the wire to the central
site. Note that - in contrast to the RF or Raman couplings along the wires -
all these terms are off-resonant because the drive frequency $\omega
=\epsilon_{1}+\epsilon_{2}$ is detuned from from the
corresponding transition frequencies $\epsilon_{1},\epsilon
_{2}$. Thus these terms average to zero, provided
$\Omega^{\prime},\Omega^{\prime\prime}\ll\epsilon$. We will neglect them in
the following discussion.

To illustrate the physics of pair hopping we consider now a single plaquette
$j=1,2$ (dotted box in Fig.~\ref{fig:model2}b). We can write the corresponding
Hamiltonian in a frame rotating with $\omega$ as
\begin{align*}
\tilde{H}_{\square} &  =\epsilon_2  a_{1a}^{\dagger}a_{1,a}- \epsilon_1 a_{2a}%
^{\dagger}a_{2,a}  -\left(  \Omega a_{1a}^{\dagger}a_{2a}e^{-i2\epsilon
t}+\mathrm{h.c.}\right)  \\
&  +\epsilon_2 a_{1b}^{\dagger}a_{1,b}- \epsilon_1 a_{2b}^{\dagger}a_{2b}
-\left(  \Omega a_{1b}^{\dagger}a_{2b}e^{-i2\epsilon t}+\mathrm{h.c.}\right)
\\
&  (-\epsilon^{\prime})\sum_{\sigma=\uparrow,\downarrow}a_{\sigma c}^{\dagger
}a_{\sigma c}+Ua_{\uparrow c}^{\dagger}a_{\downarrow c}^{\dagger}a_{\downarrow
c}a_{\uparrow c}\\
&  -\sum_{\alpha=a,b}t_{ma}\left(  a_{\uparrow c}^{\dagger}a_{j\alpha
}+a_{\downarrow c}^{\dagger}a_{j\alpha}+\mathrm{h.c.}\right)
\end{align*}
For the sites on the two wires we have again used the notation $a_{1\uparrow
,a}\rightarrow a_{1a}$, $a_{1\downarrow,b}\rightarrow a_{1b}$ and
$a_{2\downarrow,a}\rightarrow a_{2a}$, $a_{2\uparrow,b}\rightarrow a_{j,b}$.
The first line is again the Raman or RF hopping between sites
$2\leftrightarrow1$ with energies $-\epsilon_1$ and $+\epsilon_2$, respectively,
due to absorption (emission) of a photon $\omega=\epsilon_{1}%
+\epsilon_{2}$, which is tuned to compensate the energy
difference. The second line corresponds to the central site with interaction
and hopping. The last two lines describe a spin-flip on site $c$, and a Raman
assisted hopping and accompanying spin flip from the wires to the central site
$c$. 

Let us now analyze the various processes on the plaquette according to the
above Hamiltonian. We will argue that an adiabatic elimination of the central
site will result in an effective Hamiltonian for the two wires where single
particle tunneling is suppressed, while atoms can hop pairwise between the wires.

\begin{itemize}
\item \emph{Suppression of single particle interwire hopping.} Consider a
single atom on the plaquette which occupies initially, say, site $1a$, i.e.
has spin $\uparrow$. The particle hopping to and from the central site
preserves spin, and thus it can only tunnel along the diagonal $1\uparrow
,a\rightarrow c\uparrow\rightarrow1\uparrow,b$. In view of the energy mismatch
$+\epsilon_2$, 0,$-\epsilon_1$ this tunneling process is not
energy conserving and thus will occur only as virtual process,
which renormalizes the (single particle) tunneling parameters and onsite
shifts of the wire sites. \newline[Note: The above argument ignores the effect
of the drive on the central site. We argued above that this coupling will be
small for $\Omega^{\prime}\ll2\epsilon$, but can result in a (weak) energy
conserving transition $1\uparrow,a\rightarrow1\downarrow,b$. We can suppress
such terms by tilting the plaquette, so that the energies of the four lattice
$1a,2a,1b,1b$ sites are $+\epsilon_{1}',-\epsilon_{2}',+\epsilon_{2}'%
,-\epsilon_{1}'$ with $\epsilon_{1}'-\epsilon_{2}'\gg t_{ma},\Omega$. This makes
the spin-flip interwire tunneling an energy non-conserving process, i.e the
tunneling terms of the form $a_{ja}^{\dagger}a_{jb}$ will be absent.]\newline

\item \emph{Interwire pair hopping. }While the single particle interwire hops
$1\uparrow,a\rightarrow1\uparrow,b$ and $2\downarrow,a\rightarrow
1\downarrow,b$ are individually forbidden, the joint hopping is energetically
allowed. For this pair hopping to happen the two particles must occupy
simultaneously the central site to be able to exchange energy, i.e. to
interact. An adiabatic elimination of the central site therefore gives a term
$Wa_{1b}^{\dagger}a_{2b}^{\dagger}a_{1a}a_{2a}+\mathrm{h.c}.$, where in the
limit of large $U$%
\[
W=-\left(  \frac{1}{\epsilon_2}-\frac{1}{\epsilon_1
}\right)  ^{2}\frac{t_{am}^{2}t_{bm}^{2}}{U}.
\]

Particle assisted tunneling, where, for example, a particle hops
$1\uparrow,a\rightarrow2\uparrow,b$ while a second particle $2\downarrow
,a\rightarrow2\downarrow,a$ remains on wire $a$, is suppressed by energy
conservation. A process $1\uparrow,a\rightarrow2\uparrow,b$ while
$2\downarrow,a\rightarrow1\uparrow,a$ requires a spinflip on the central site
which we argued to be small; in addition this process can be suppressed by
tilting the plaquette.

\item \emph{Parity-preserving perturbations.} Imperfections which preserve the
parity symmetry are also present. Local off-sets of the form:
\begin{align}
H_{\text{off}} &  =\sum_{j\text{odd}}\xi_{j\uparrow,a}a_{j\uparrow,a}%
^{\dagger}a_{j\uparrow,a}+\sum_{j\text{even}}\xi_{j\downarrow,a}%
a_{j\downarrow,a}^{\dagger}a_{j\downarrow,a}+\nonumber\\
&  +\sum_{j\text{even}}\xi_{j\uparrow,b}a_{j\uparrow,b}^{\dagger}%
a_{j\uparrow,b}+\sum_{j\text{odd}}\xi_{j\downarrow,b}a_{j\downarrow
,b}^{\dagger}a_{j\downarrow,b}%
\end{align}
are also generated within second order perturbation theory, the corresponding
coefficients being:
\begin{align}
\xi_{j\uparrow,a}&=\frac{t_{am}^{2}}{\epsilon_2}, \;
\xi_{j\uparrow,b}=-\frac{t_{bm}^{2}}{\epsilon_1}, \nonumber\\
\xi_{j\downarrow,a}&=\frac{t_{am}^{2}}{\epsilon_2},\;
\xi_{j\downarrow,b}=\frac{t_{bm}^{2}}{\epsilon_2}
\end{align}
Moreover, in fourth order perturbation theory, additional diagonal interactions
emerge, induced by virtual processes where two particles from the site
$j,j^{\prime}$ belonging to the same plaquette hop into the intermediate
sites, and subsequently hop back to $j,j^{\prime}$. The corresponding terms
read:
\[
H_{\text{diag}}=K\sum_{j,j^{\prime}}\sum_{\sigma,\sigma^{\prime}%
=\uparrow,\downarrow}\sum_{\alpha,\alpha^{\prime}=a,b}n_{j\sigma,\alpha
}n_{j^{\prime}\sigma^{\prime},\alpha^{\prime}}%
\]
with coefficients:
\begin{equation}
K\simeq\left(  \frac{1}{\epsilon_2}-\frac{1}{\epsilon_1
}\right)  ^{2}\frac{t_{am}^{2}t_{bm}^{2}}{U},
\end{equation}
are also present. Regarding the relevance of such additional terms in the
effective Hamiltonian, we refer the reader to the discussion of the
imperfections in the following section.
\end{itemize}

\subsection{Numerical analysis of a single plaquette}
\begin{figure}[t]
\includegraphics[width = 0.98 \columnwidth]{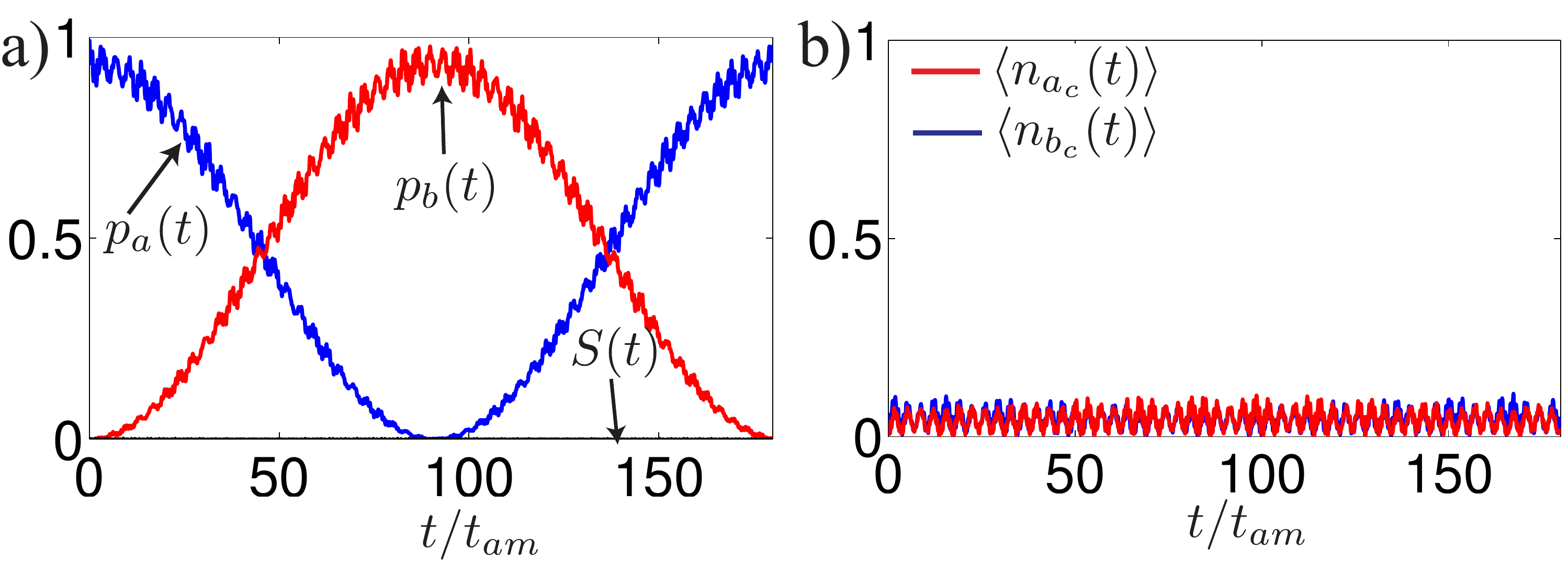}
\caption{Real time evolution for the parameters in the main text. a) Time evolution of the expectation values $p_a(t) = \langle n_{a,1}(t) n_{a,2}(t)\rangle$ and $p_b(t) = \langle n_{b,1}(t) n_{b,2}(t)\rangle$ that indicate the pair hopping. Single particle hopping indicated by $S(t) = \sum_{i,j = 1,2} \langle n_{a,i}(t) n_{b,j}(t)\rangle$ is of the order of $10^{-2}$. b) The  occupation of the intermediate sites $\langle n_{a_c}\rangle, \langle n_{b_c}\rangle$, is very small \label{fig:Energy_conservation}}
\end{figure}

In the following we investigate a single plaquette and carry out a real-time evolution under the Hamiltonian $\tilde{H}_{\square} $
starting from a state with two particles on the upper wire,
$|\Psi\rangle= a^{\dagger}_{1} a_{2}^{\dagger}|0\rangle$. We plot, as a function of time, the double occupancy $p_{a}(t) = \langle
n_{a,1}(t) n_{a,2}(t)\rangle$ and $p_{b}(t) = \langle n_{b,1}(t)n_{b,2}(t)\rangle$ in the upper resp. lower wire. A large value of $p_{b}(t)$ is an indication of pair hopping. Single particle hopping, which is indicated via a finite value of $S(t) = \sum_{i,j = 1,2} \langle
n_{a,i}(t) n_{b,j}(t)\rangle$ should be suppressed. Taking $t_{am} = t_{bm} = 1$ as the unit of energy, we depict, in Fig.~\ref{fig:Energy_conservation}a), numerical results for $p_{a,b}(t)$ and $S(t)$, where $\epsilon_2 = 2\epsilon_1 = 2$, $U=-20$, $\epsilon' = 1/2(U - (\epsilon_2 - \epsilon_1)) + \delta$, and $\delta = 1$. We find a large and finite value for the pair hopping, while the single particle hopping is of the order of $10^{-2}$. The occupation of the central sites (Fig.~\ref{fig:Energy_conservation}b) is very small.

\subsection{Relation to Model 2}

As a final remark we will relate the present model to the one described in the
following subsection, where the role of tunneling and Raman beams is
interchanged. We can make a time-dependent transformation to rewrite the above
plaquette Hamiltonian as%

\begin{align*}
\tilde{H}_{\square} &  =-\left(  \Omega a_{2a}^{\dagger}a_{1a}+\mathrm{h.c.}%
\right)  -\left(  \Omega a_{1b}^{\dagger}a_{2b}+\mathrm{h.c.}\right)  \\
&  +Ua_{\uparrow c}^{\dagger}a_{\downarrow c}^{\dagger}a_{\downarrow
c}a_{\uparrow c}\\
&  -t_{ma}\left(  a_{\uparrow c}^{\dagger}a_{1a}e^{-i(\epsilon+\epsilon
^{\prime})t}+a_{\downarrow c}^{\dagger}a_{2a}e^{i(\epsilon-\epsilon^{\prime
})t}+\mathrm{h.c.}\right)  \\
&  -t_{mb}\left(  a_{\downarrow c}^{\dagger}a_{1b}e^{+i(\epsilon
-\epsilon^{\prime})t}+a_{\uparrow c}^{\dagger}a_{2b}e^{-i(\epsilon
+\epsilon^{\prime})t}+\mathrm{h.c.}\right)  .
\end{align*}
The first line now looks like a tunneling Hamiltonian, while the last two
lines correspond to time-dependent hoppings to the central site with
frequencies $\epsilon\pm\epsilon^{\prime}$. In this rewriting of the model the
lattice offsets have been converted to effective time-dependent RF or Raman
couplings. In the following section we investigate a model, which is a
generalization of this scheme. 

\section{SI: Effective pair-hopping Hamiltonian - Model 2}




\subsection{Building block: Single wire with resonant coupling to intermediate
states}

\begin{figure}[t]
\includegraphics[width = 0.35\columnwidth]{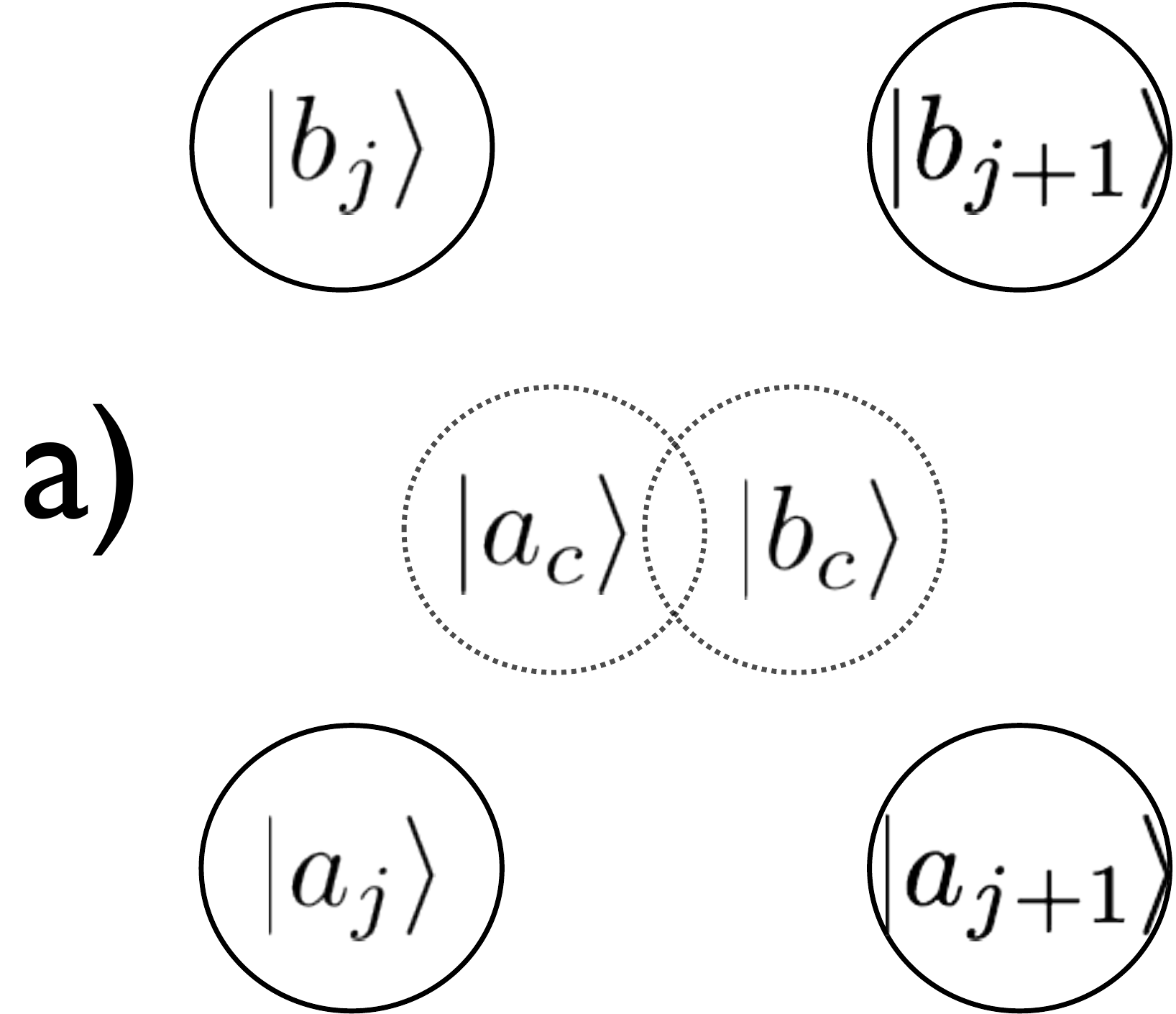} \smallskip
\vspace*{0.55cm} \smallskip
\includegraphics[width = 0.85\columnwidth]{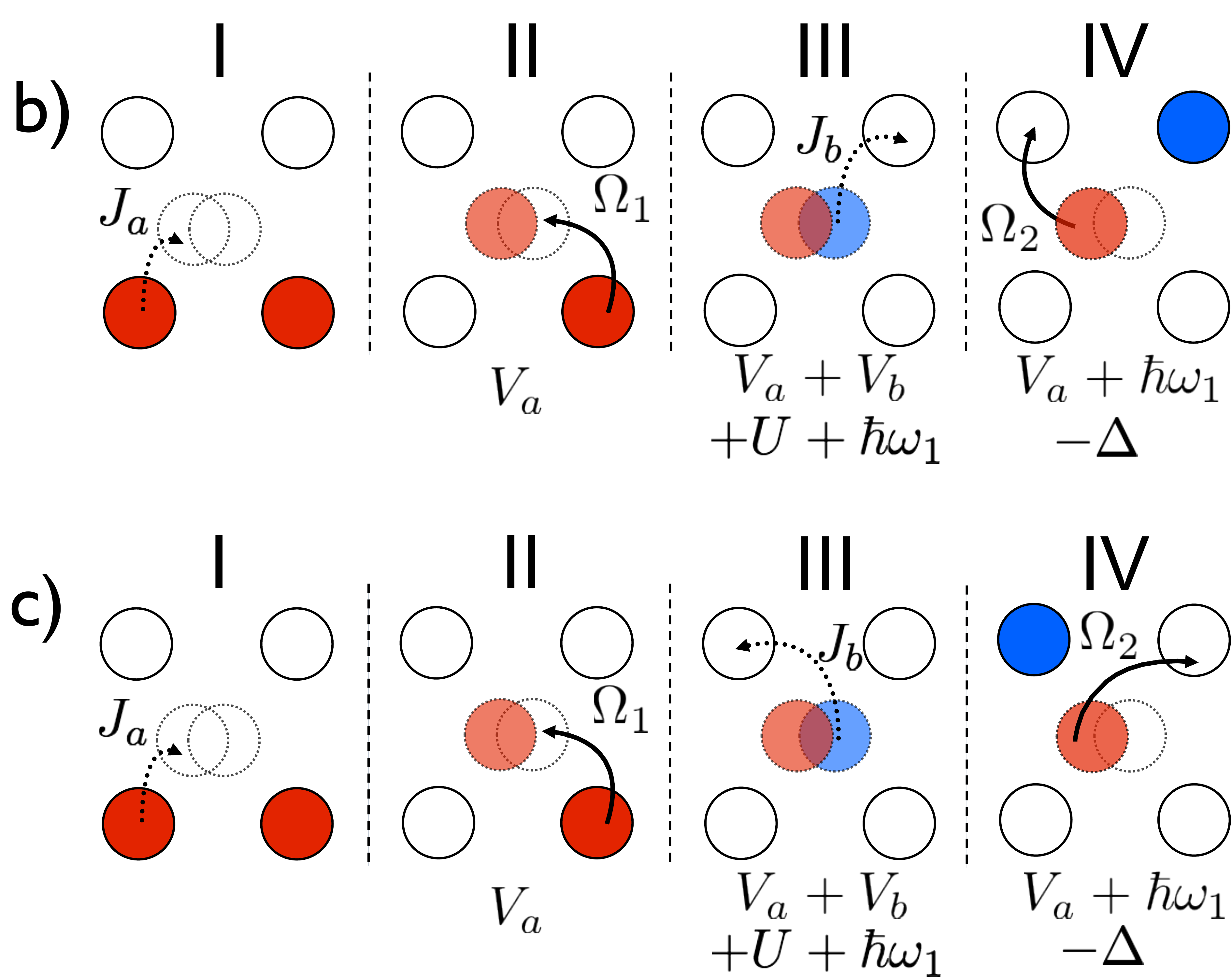}
\includegraphics[width = 0.75\columnwidth]{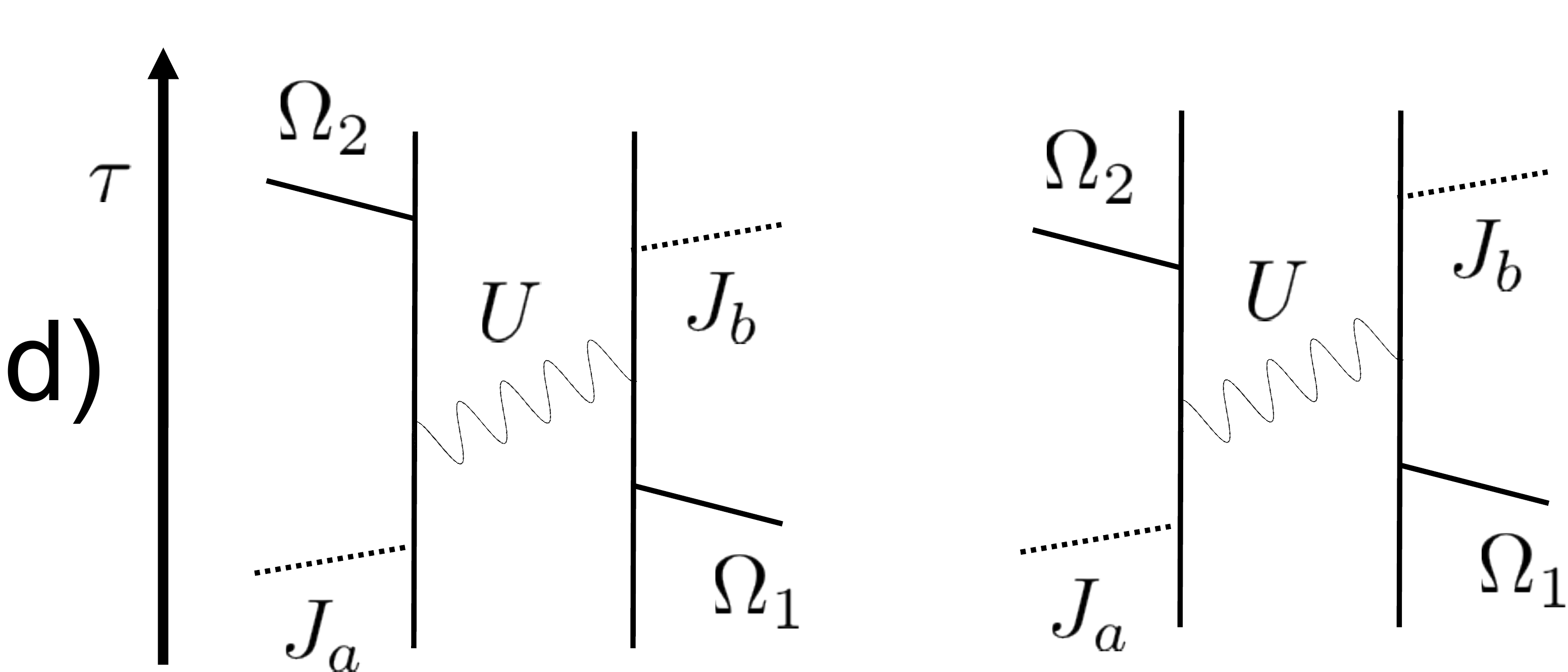} \caption{Panel
\textit{a)}: Schematic Hilbert space of a single plaquette, including states
in the upper and lower wires ($|a_{j/j+1}\rangle,|b_{j/j+1}\rangle$) and
states in the center of the plaquette $|a\rangle_{c},|b\rangle_{c}$. Panels
\textit{b,c)}: Schematic examples of perturbation theory processes generating
the pair tunneling term. Below each perturbation step, denoted as I, II, III
and IV, the energy of the intermediate state is indicated. In case of constant
$\Omega_{1/2,j}$, these two processes interfere destructively. Panel
\textit{d)}: Diagrams corresponding to the illustrations in panels \textit{b}
and \textit{c)}, where $\tau$ defined the time evolution of the perturbative
steps, and dashed (thin) lines denote tunneling (Raman) processes. }%
\label{fig:diagrams}%
\end{figure}

The basic building block of model II is represented by a
single plaquette, illustrated in Fig.~\ref{fig:diagrams}a, where the upper
sites belong to the $a$-wire and the lower sites to the $b$- one, while the
middle sites are employed as intermediate sites (which will then be
adiabatically eliminated) to generate the pair tunneling term. Before
discussing the emergence of the full pair-tunneling Hamiltonian, we show how
one can couple one of the wires to the intermediate sites such that pair
tunneling from the wire to the middle sites is generated in second order
perturbation theory. This discussion will then be naturally extended to the
full plaquette treatment, where the emergence of inter-wire pair tunneling
will emerge as a combination of a pair tunneling process from the $a$-wire to
the central sites, and from the central sites to the $b$-wire.

We start by analyzing the system composed by the lower part of the plaquette
in Fig.~\ref{fig:diagrams}a. Atoms in the atomic state $|a\rangle$ can occupy
the 'wire' sites $|a_{j}\rangle, |a_{j+1}\rangle$ and the central site
$|a_{c}\rangle$; atoms in the atomic state $|b\rangle$ can occupy the central
site $|b_{c}\rangle$. The microscopic Hamiltonian $\tilde{H}$ for the
subsystem can be decomposed as a sum of three terms:
\begin{equation}
\tilde{H}_{a}=\tilde{H}_{t,a}+\tilde{H}_{C}+\tilde{H}_{\Omega,a}.
\end{equation}
The first term describes tunneling of the fermions within the wire:
\begin{equation}
\tilde{H}_{t,a}=-\tilde{t}_{a}(a^{\dagger}_{j}a_{j+1}+\text{h.c.}),
\end{equation}
where $\tilde{t}_{a}$ is the tunneling amplitude, and $a^{\dagger}_{j}
(a_{j})$ are creation (annihilation) operators of fermions in the state
$|a_{j}\rangle$. The second term
\begin{align}
\tilde{H}_{C}  &  =V_{a}a^{\dagger}_{c}a_{c}+V_{b}b^{\dagger}_{c}%
b_{c}+Ua^{\dagger}_{c}a_{c}b^{\dagger}_{c} b_{c},
\end{align}
describes the two intermediate states $|a_{c}\rangle$ and $|b_{c}\rangle$ in
the middle of the plaquette. Here, $a_{c}^{\dagger}(a_{c})$ and $b_{c}%
^{\dagger}(b_{c})$ are creation (annihilation) operators of fermions for the
intermediate states $|a_{c}\rangle$ and $|b_{c}\rangle$, respectively, $V_{a}$
and $V_{b}$ are the corresponding potential off-sets, and $U$ is the
interparticle interaction. The last term
\begin{align}
\tilde{H}_{\Omega,a}  &  =J_{a}(a^{\dagger}_{c}(a_{j}+a_{j+1})+\text{h.c.}%
)+\nonumber\\
&  -\hbar(b^{\dagger}_{c} (\Omega_{1,j}a_{j}+\Omega_{1,j}a_{j+1}%
)e^{-i\omega_{1} t}+\text{h.c.})
\end{align}
describes the coupling between the wires and the states in the middle of the
plaquette: Atoms in the $|a\rangle$ state can tunnel from the wire to the
intermediate site $|a_{c}\rangle$, with a tunneling coefficient $J_{a}$, and
can be transferred to the state $|b_{c}\rangle$ via a Raman process
characterized by (space-dependent) Rabi frequency $\Omega_{1, j}$ and detuning
$\hbar\omega_{1}$. We work here with time-dependent fields assuming a rotating
wave approximation ($\Omega_{1,j}\ll\omega_{1}$). Note that the same results
can be obtained by describing Raman process as an auxiliary quantized single
photon mode.

We are now interested in the dynamics of such basic building block in the
regime where single particle occupation in the intermediate sites is
suppressed as a far-off resonant state, while double occupancies of the form
$a^{\dagger}_{c}b^{\dagger}_{c}|0\rangle$ are allowed (here, $|0\rangle$ is
the fermionic vacuum). In order to illustrate it, we perform a quasi-degenerate
perturbation theory (within the rotating wave approximation) in the limit
$|U|, \hbar\omega_{1},V_{a}, V_{v}\gg\tilde{t}_{a}, |\hbar\Omega_{1,j/j+1}|$,
with the additional quasi-resonant condition:
\begin{equation}
\label{deltaU}U=-V_{a}-V_{b}-\hbar\omega_{1}+\delta_{U}, \quad|\delta
_{U}|\simeq\tilde{t}_{a},|\hbar\Omega_{1}|.
\end{equation}
In this limit, the two states $a^{\dagger}_{j}a^{\dagger}_{j+1}|0\rangle$ and
$a^{\dagger}_{c}b^{\dagger}_{c}|0\rangle$ are the only quasi-degenerate with
total number of fermionic particles equal to two. After eliminating the states
with singly occupied intermediate sites in the second order perturbation
theory, we obtain the following effective Hamiltonian:
\begin{align}
H_{a}^{\text{eff},2}  &  =-t_{a}(a^{\dagger}_{j}a_{j+1}+\text{h.c.}%
)+W_{a}(a^{\dagger}_{c}b^{\dagger}_{c}a_{j}a_{j+1}+ \text{h.c.})+\nonumber\\
&  +\delta_{U} (a^{\dagger}_{c}a_{c}b^{\dagger}_{c}b_{c})+\xi_{a,j}a^{\dagger
}_{j}a_{j}+\xi_{a,j+1}a^{\dagger}_{j+1}a_{j+1},%
\end{align}
with a renormalized tunneling rate
\begin{align}
t_{a}=\tilde{t}_{a}-(J_{a})^{2}/V_{a}-\Omega_{1,j+1} \Omega_{1,j}^{*}%
/(V_{a}-\hbar\omega_{1}),
\end{align}
an effective interaction in the intermediate site $\delta_{U}$, and a
potential off-set on the a-wire:
\begin{equation}
\label{xia}\xi_{a,j}=-\frac{J_{a}^{2}}{V_{a}}-\frac{|\Omega_{a,j}|^{2}}%
{(V_{b}+\hbar\omega_{1})}.
\end{equation}
Moreover, a pair tunneling from the wire to the central sites emerges, with
coefficient:
\begin{align}
W_{a}=\left(  \frac{1}{V_{a}}+\frac{1}{V_{b}+\hbar\omega_{1}}\right)  \hbar
J_{a}(\Omega_{1,j+1}-\Omega_{1,j}).
\end{align}
Note that in order to have a finite coefficient of the pair tunneling,
space-dependent Rabi frequencies are required. This is due to the fact that,
for a short-range interparticle interaction, one has to use spatially
inhomogeneous Rabi frequencies in order to change the symmetry of the spatial
part of the wave function for two particles from antisymmetric (for two
initial particle on one of the wires) to symmetric (for two particles in the
center of a plaquette).

In a similar way, one describe the system of the $b$-wire (upper part of the
plaquette) coupled to the intermediate sites. The corresponding microscopic
Hamiltonian reads:
\begin{equation}
\tilde{H}_{b}=\tilde{H}_{t, \Delta,b}+\tilde{H}_{C}+\tilde{H}_{\Omega,b},
\end{equation}
where
\begin{align}
\label{Ht}\tilde{H}_{t,\Delta,b}=-\tilde{t}_{b}(b^{\dagger}_{j}b_{j+1}%
+\text{h.c.})- \Delta(b^{\dagger}_{j}b_{j} + b^{\dagger}_{j+1}b_{j+1}),
\end{align}
describes tunneling in the $b$-wire, and an off-set $\Delta$. Here,
$b^{\dagger}_{j}(b_{j})$ are creation/annihilation operators corresponding to
the states $|b_{j}\rangle$. The term
\begin{align}
\label{HOmega}\tilde{H}_{\Omega,b}  &  =-\hbar(a^{\dagger}_{c} (\Omega
_{2,j}b_{j}+\Omega_{2,j+1}b_{j+1})e^{i\omega_{2}t } +\text{h.c.})+\nonumber\\
&  +J_{b}(b^{\dagger}_{c}(b_{j}+b_{j+1})+\text{h.c.}),
\end{align}
describes the tunneling between the $b$-wire and the intermediate site
$|b_{c}\rangle$ with the amplitude $J_{b}$, and the coupling between the
$b$-wire and the intermediate site $|a_{c}\rangle$ via a Raman process with
detuning $\hbar\omega_{2}$ and Rabi frequency $\Omega_{2,j}$.

The energy off-set is introduced in order to suppress single particle
tunneling between the wires (as discussed below). This off-set, however, has
to match the condition
\begin{equation}
\label{delta_1}2\Delta=\hbar(\omega_{1}+\omega_{2})
\end{equation}
in order to ensure resonant pair-tunneling.

After eliminating the states with singly occupied intermediate sites in the
second order perturbation theory in $|U|,\hbar\omega_{2}, V_{a}, V_{b}\gg t,
\hbar\Omega$, we obtain the effective Hamiltonian:
\begin{align}
H_{b}^{\text{eff},2}  &  =-t_{b}(b^{\dagger}_{j}b_{j+1}+\text{h.c.}%
)+W_{b}(a^{\dagger}_{c}b^{\dagger}_{c}b_{j}b_{j+1}+ \text{h.c.})+\nonumber\\
&  +\delta_{U} (a^{\dagger}_{c}a_{c}b^{\dagger}_{c}b_{c})+\xi_{b,j}b^{\dagger
}_{j}b_{j}+\xi_{b,j+1}b^{\dagger}_{j+1}b_{j+1}%
\end{align}
with
\begin{align}
W_{b}=\left(  \frac{1}{V_{b}+\Delta}+\frac{1}{V_{a}-\hbar\omega_{2}+\Delta
}\right)  \hbar J_{b}(\Omega_{2,j+1}-\Omega_{2,j}),\nonumber
\end{align}
\begin{equation}
\label{tb}t_{b}=\tilde{t}_{b}-(J_{b})^{2}/(V_{b}+\Delta) -\Omega_{2,j+1}%
\Omega_{2,j}^{*}/(V_{a}-\hbar\omega_{2}+\Delta),
\end{equation}
\begin{equation}
\label{xib}\xi_{b,j}=-\frac{J_{b}^{2}}{(V_{b}+\Delta)}-\frac{|\Omega
_{b,j}|^{2}}{(V_{a}+\Delta+\hbar\omega_{2})}.
\end{equation}

Despite its simplicity, this setup already displays the fundamental features
of the emergence of inter-wire pair tunneling. We will now show
how the two wires can be resonantly coupled to the intermediate sites while
avoiding single particle inter-wire tunneling.

\subsection{Coupled wires}

\begin{figure}[t]
\includegraphics[width = 0.65\columnwidth]{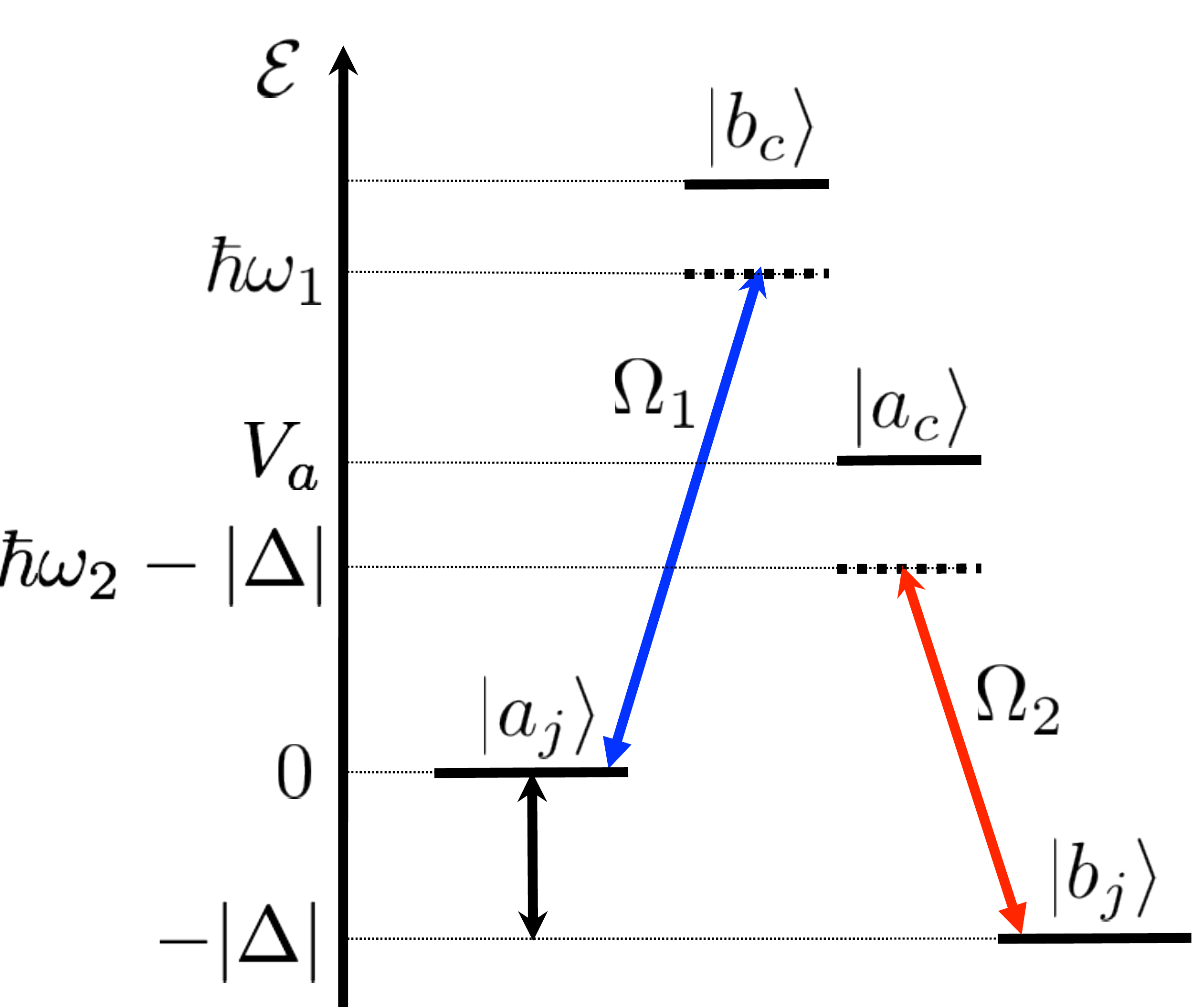} \caption{Single
particle level scheme, illustrating how the off-resonant condition between
$|a\rangle_{j}$ and $|b\rangle_{j}$ can be engineered by fulfilling
Eq.~\eqref{delta_2}.}%
\label{fig:level1}%
\end{figure}

Our goal is to derive an effective Hamiltonian for the full plaquette where
\textit{i)} single occupancies in the intermediate sites are suppressed, and
\textit{ii)} single particle tunneling between the wires is suppressed by
energy constraints. The first condition is fulfilled in the regime discussed in
the previous section; moreover, as single particle tunneling between the wires
is always driven by a single auxiliary field~\cite{note2}, the condition:
\begin{equation}
\label{delta_2}|\Delta-\hbar\omega_{1}|, |\Delta-\hbar\omega_{2}|\gg\tilde
{t}_{a}, \tilde{t}_{b}, |\hbar\Omega_{1,j/j+1}|, |\hbar\Omega_{2,j/j+1}|
\end{equation}
guarantees that inter-wire single particle tunneling are far-off resonant
processes, and thus suppressed within perturbation theory (see
Fig.\ref{fig:level1}).

In the perturbative regime $|U|, \hbar\omega_{1}, \hbar\omega_{2}, \Delta,
V_{a}, V_{v}\gg\tilde{t}_{a}, \tilde{t}_{b}, |\hbar\Omega_{1,j/j+1}|,
|\hbar\Omega_{2,j/j+1}|$, after combining the results of the previous section,
one obtains the following effective Hamiltonian:
\begin{align}
\label{H2eff}H^{\text{eff},2}  &  =-t_{a}(a^{\dagger}_{j}a_{j+1}%
+\text{h.c.})-t_{b}(b^{\dagger}_{j}b_{j+1}+\text{h.c.})+\nonumber\\
&  +W_{a}(a^{\dagger}_{c}b^{\dagger}_{c}a_{j}a_{j+1}+ \text{h.c.}%
)+W_{b}(a^{\dagger}_{c}b^{\dagger}_{c}b_{j}b_{j+1}+ \text{h.c.})+\nonumber\\
&  +\delta_{P} (a^{\dagger}_{c}a_{c}b^{\dagger}_{c}b_{c})+\xi_{a,j}a^{\dagger
}_{j}a_{j}+\xi_{a,j+1}a^{\dagger}_{j+1}a_{j+1}+\nonumber\\
&  +\xi_{b,j}b^{\dagger}_{j}b_{j}+\xi_{b,j+1}b^{\dagger}_{j+1}b_{j+1}.
\end{align}
Here, the first line describes renormalized intrawire tunneling; the second
one contains terms coupling the $a$ and $b$-wire to the intermediate sites,
respectively. Finally, in analogy with the previous discussion, the last two
lines describe a residual interaction in the intermediate sites due to the
quasi-resonant condition in Eq.~\eqref{deltaU}, and local potential offsets.
Beyond constituting the basis for the realization of the pair-tunneling
Hamiltonian, Eq.~\eqref{H2eff} displays interesting physics by itself, as its
effective description in terms of two Luttinger liquids coupled with a 1D
superconductor (as an effective description of the intermediate wire composed
by the center sites of each plaquette) has been also investigated in the
context of emergent Majorana edge states~\cite{fidkowski2011bib}.

\begin{figure}[t]
\includegraphics[width = 0.65\columnwidth]{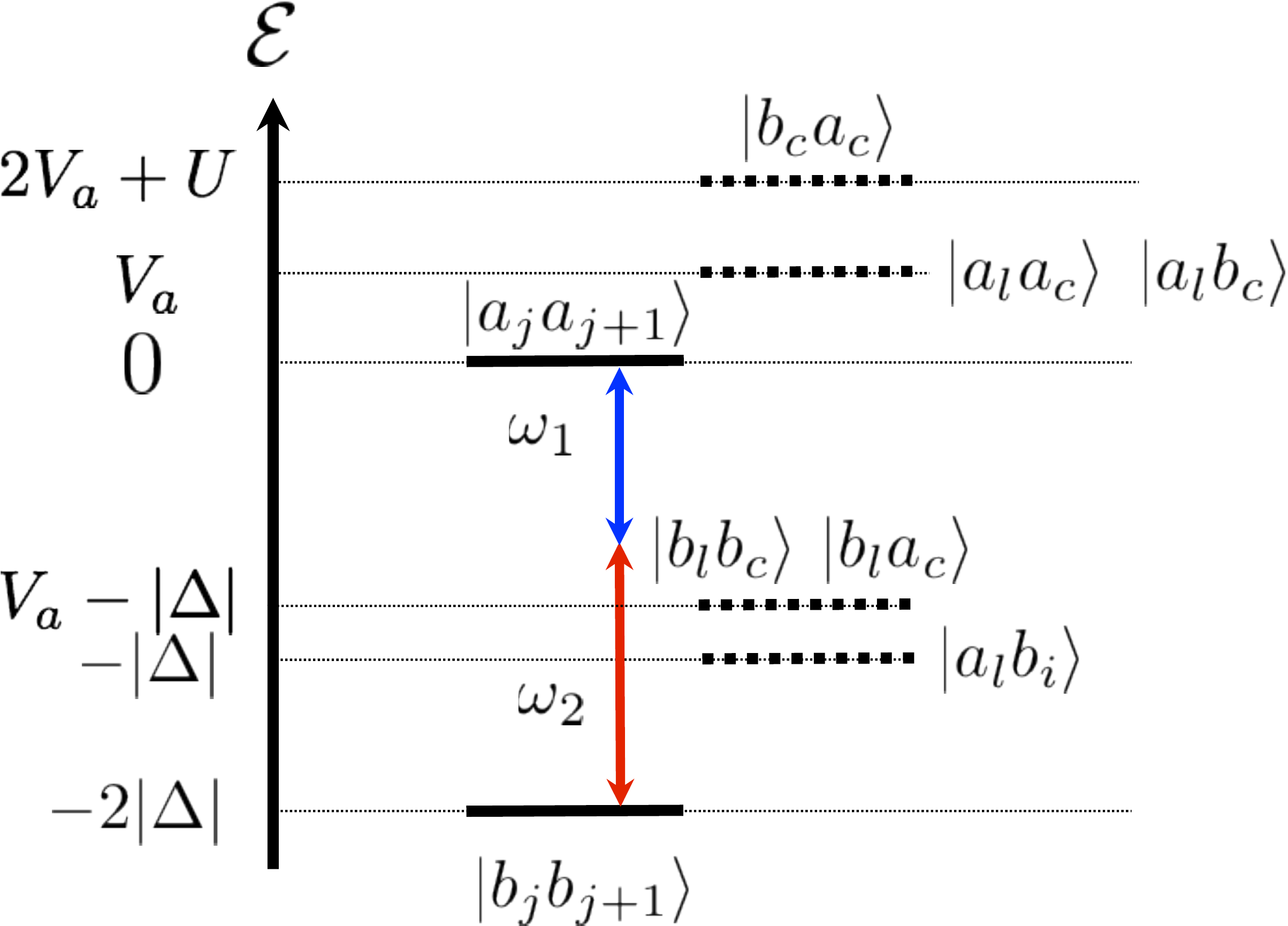}
\includegraphics[width = 0.65\columnwidth]{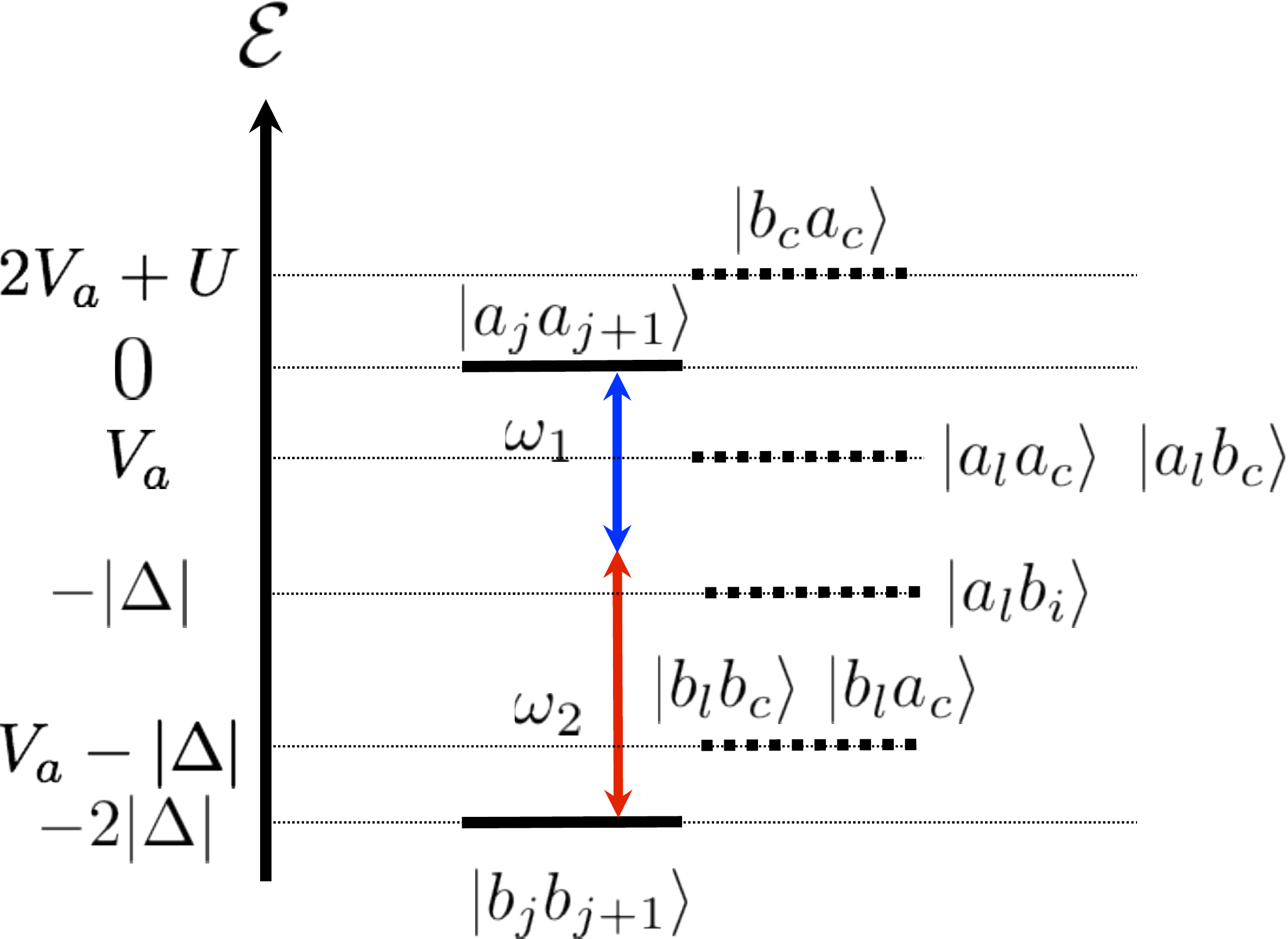} \caption{Schematic
two-particle level scheme in a single plaquette (here, $V_{a}=V_{b}>0$ in the
upper panel, and $V_{a}=V_{b}<0$ in the lower, for the sake of clarity). The
indices $l,i$ run over $(j, j+1)$, indicating the multiplicity of the various
intermediate states (15 states are involved). The Raman detunings $\hbar
\omega_{1/2}$ are indicated by the green/red line, respectively, showing the
two particle resonant condition in Eq.~\eqref{delta_1}. }%
\label{fig:level2}%
\end{figure}

\subsection{Inter-wire pair tunneling}

\subsubsection{Adiabatic elimination of the intermediate sites}

The pair tunneling Hamiltonian discussed in the text is then obtained by
adiabatically eliminating the central sites from Eq.~\eqref{H2eff} in the
regime $\delta U\gg t, \hbar\Omega$, where the quasi-resonant condition in
Eq.~\eqref{deltaU} is not met anymore. The effective Hamiltonian will then
read:
\begin{align}
\label{H4eff}H^{\text{eff}, 4}  &  =-t_{a}(a^{\dagger}_{j}a_{j+1}%
+\text{h.c.})-t_{b}(b^{\dagger}_{j}b_{j+1}+\text{h.c.})+\nonumber\\
&  +W(b^{\dagger}_{j}b^{\dagger}_{j+1}a_{j}a_{j+1}+ \text{h.c.})+\nonumber\\
&  +\xi_{a,j}a^{\dagger}_{j}a_{j}+\xi_{a,j+1}a^{\dagger}_{j+1}a_{j+1}%
+\nonumber\\
&  +\xi_{b,j}b^{\dagger}_{j}b_{j}+\xi_{b,j+1}b^{\dagger}_{j+1}b_{j+1}%
+H^{\text{imp}},%
\end{align}
where the second line describe the pair tunneling term between the wires
(which, in perturbation theory, emerges as a sum of terms similar to the ones
illustrated in Fig.~\ref{fig:diagrams}), with coefficient:
\begin{align}
\label{H_W}W  &  =\left(  \frac{1}{V_{a}}+\frac{1}{V_{b}+\hbar\omega_{1}%
}\right)  \left(  \frac{4\hbar^{2}J_{a}J_{b}}{\delta U}\right)  \times
\nonumber\\
&  \times\left(  \frac{1}{\Delta+V_{a}-\hbar\omega_{2}}+ \frac{1}{\Delta
+V_{b}}\right)  \times\nonumber\\
&  \times(\Omega_{1,j+1}-\Omega_{1,j})(\Omega_{2,j}^{*}-\Omega_{2,j+1}^{*}),
\end{align}
while $H^{\text{imp}}$ denotes additional imperfections (such as intra-wire
and inter-wire interactions) emerging from additional fourth order processes.
While all of them preserve parity symmetry, some of them may change the energy
splitting between the two degenerate states in the topological region
discussed in the text from exponential to power law as a function of the
system size.

\subsubsection{Additional terms}

Let us now discuss the imperfections described by $H^{\text{imp}}$.

\paragraph{Intra-wire interactions.}

These interactions are generated via virtual processes, where, e.g., two
particles on nearest-neighbor sites in the $a$-wire are transferred to the
intermediate sites $|a\rangle_{c}$ and $|b\rangle_{c}$ via tunneling and Raman
assisted process, respectively, and then move back to the same two sites. The
effective contribution in the Hamiltonian reads:
\begin{align}
G_{a}n_{a,j}n_{a,j+1}+G_{b}n_{b,j}n_{b,j+1}%
\end{align}
and the corresponding pre-factors in fourth-order perturbation theory are:
\begin{align}
G_{a}  &  =-\left(  \frac{1}{V_{a}}+\frac{1}{\hbar\omega_{1}+V_{b}}\right)
^{2} \frac{(\hbar J_{a})^{2}}{\delta U}\times\\
&  \times(|\Omega_{1,j}|^{2}+|\Omega_{1,j+1}|^{2}-\Omega_{1,j}^{*}%
\Omega_{1,j+1}-\Omega_{1,j+1}^{*}\Omega_{1,j}),\nonumber
\end{align}
and
\begin{align}
G_{b}  &  =-\left(  \frac{1}{\Delta+V_{b}}+\frac{1}{-\hbar\omega_{2}%
+V_{a}+\Delta}\right)  ^{2}\frac{(\hbar J_{b})^{2}}{\delta U}\times\\
&  \times(|\Omega_{2,j}|^{2}+|\Omega_{2,j+1}|^{2}-\Omega_{2,j}^{*}%
\Omega_{2,j+1}-\Omega_{2,j+1}^{*}\Omega_{2,j}).\nonumber
\end{align}
The effect of these terms is to renormalized the single wire Luttinger
parameter, producing only some quantitative shifts in the model Hamiltonian
phase diagram. In case $G_{b},G_{a}<0$, the topological phase is expected to
emerge at smaller values of $W$, as the value of $K_{A}$ becomes larger when
an additional intra-wire attraction is introduced~\cite{giamarchi_bookbib}.

\paragraph{Inter-wire interactions.}

Inter-wire terms emerge as well in fourth order perturbation theory. They are
induced by virtual processes, where a particle hops from $|a\rangle_{i}$ to
$|a\rangle_{c}$ and another one hops from $|b\rangle_{l}$ to $|b\rangle_{c}$
(or similar processes induced by the Raman couplings), and then both of them
hop back to the original wires in $|a\rangle_{k}$ and $|b\rangle_{r}$. The
contribution in the effective Hamiltonian then reads:
\begin{align}
\sum_{i,l,k,r=j,j+1}(K_{i,l,k,r} a^{\dagger}_{i}a_{l}b^{\dagger}_{k}%
b_{r}+\text{h.c.}).
\end{align}
Here, the indices $i, l, k, r$ belong to the same plaquette, that is, $i, l,
k, r \in\{j, j+1\}$. The corresponding coefficients are:
\begin{align}
K_{i,l,k,r}  &  =-\left[  \left(  \frac{1}{V_{a}}+\frac{1}{V_{b}+\Delta}
\right)  ^{2} \left(  \frac{J_{a}^{2}J_{b}^{2}}{V_{b}+V_{a}+\Delta+U} \right)
\right.  +\nonumber\\
&  +\left.  \left(  \frac{1}{V_{a}-\hbar\omega_{2}+\Delta}+\frac{1}%
{V_{b}+\hbar\omega_{1}} \right)  ^{2} \right.  \times\nonumber\\
&  \times\left.  \left(  \frac{\hbar^{4}\Omega_{1,l}\Omega_{2,r}\Omega
_{1,i}^{*}\Omega_{2,k}^{*}}{V_{b}+V_{a}+\Delta+U+\hbar\omega_{1}-\hbar
\omega_{2}} \right)  \right]  .
\end{align}
While not breaking the parity symmetry, these terms may lift the splitting of
the ground and first excited state from exponential to algebraic as a function
of the system size~\cite{fidkowski2011bib}; as such, it'd be preferable to
strongly reduce their effects in order to guarantee observability of the
topological features even in small size systems. In general, the ration
between $W$ and $K_{i,l,k, r}$ can be minimized by tuning the Hamiltonian
parameter in such a way that the sum of all processes involved in
$K_{i,l,k,r}$ is annihilated via quantum interference; this requires matching
conditions involving both interactions, off-sets and detunings. A simpler,
alternative way is to tune the potential offsets such that $V_{a}%
=-\Delta-V_{b}$,
so that the first contribution is killed via interference, and then take
$J\simeq4\Omega$; this way, the ratio $W/K_{i,l,k, r}$ will be of order
$\simeq1/(4*(16))= 1/64$, thus negligible with respect to other possible
imperfections in the system.

\paragraph{Other contributions.}

Within the effective Hamiltonian, the additional potential off-sets should
have the same effects as the intra-wire terms, and it can be shown that they
strength can be tuned by slightly shifting the initial potential off-set
$\Delta$ from exact degeneracy according to Eq.~\eqref{delta_1}. Moreover, it
can be shown (by employing a floquet formalism) that imperfect addressing of
the individual sites by the Raman beams does not qualitatively modify the
effective Hamiltonian, as the quasi-degenerate subspace within perturbation
theory is not affected.

\subsection{Numerical results}
In the following we present quantitative results on model I defined on a single plaquette. Starting from a state with two particles on the upper wire, $|\Psi\rangle = a^\dagger_1 a_2^\dagger |0\rangle$, we carry out a real time evolution under the Hamiltonian $\tilde H$. Based on our numerics, we answer the following two questions: (1) Can we suppress single particle hopping between the two wires? (2) Do we get pair hopping? Further, we check the validity of our perturbation theory assumptions. First, we investigate the effect of giving up the space dependence of the Rabi frequencies and take $\Omega_j = \Omega_{j+1}$. We expect that the pair hopping disappears. Second, we check the resonance condition for pair hopping, $\Delta = \tfrac{\hbar}{2}(\omega_1 + \omega_2)$, by comparing to a protocol where $\Delta = \hbar \omega_1$, i.e. single particle hopping is resonant. For the numerical investigation, we use the experimentally realistic parameters  $t_a = t_b = 10 Hz$, $J_a  = J_b = |\Omega_j| = 1 kHz$,  $V_b = -6 kHz$, $\hbar \omega_1 = 20 kHz$, $\hbar \omega_2 = 12 kHz$. The resonance conditions imply that 
\begin{eqnarray}
V_a = - \Delta - V_b + \delta V,\\
U = -\hbar \omega_1 - V_a - V_b + \delta U,
\end{eqnarray}
where we have defined $\delta U$ and $\Delta V$ to account for small deviations from the resonance conditions, as the are expected in a realistic experimental setup. 

\begin{figure}[t]
\includegraphics[width = 0.98 \columnwidth]{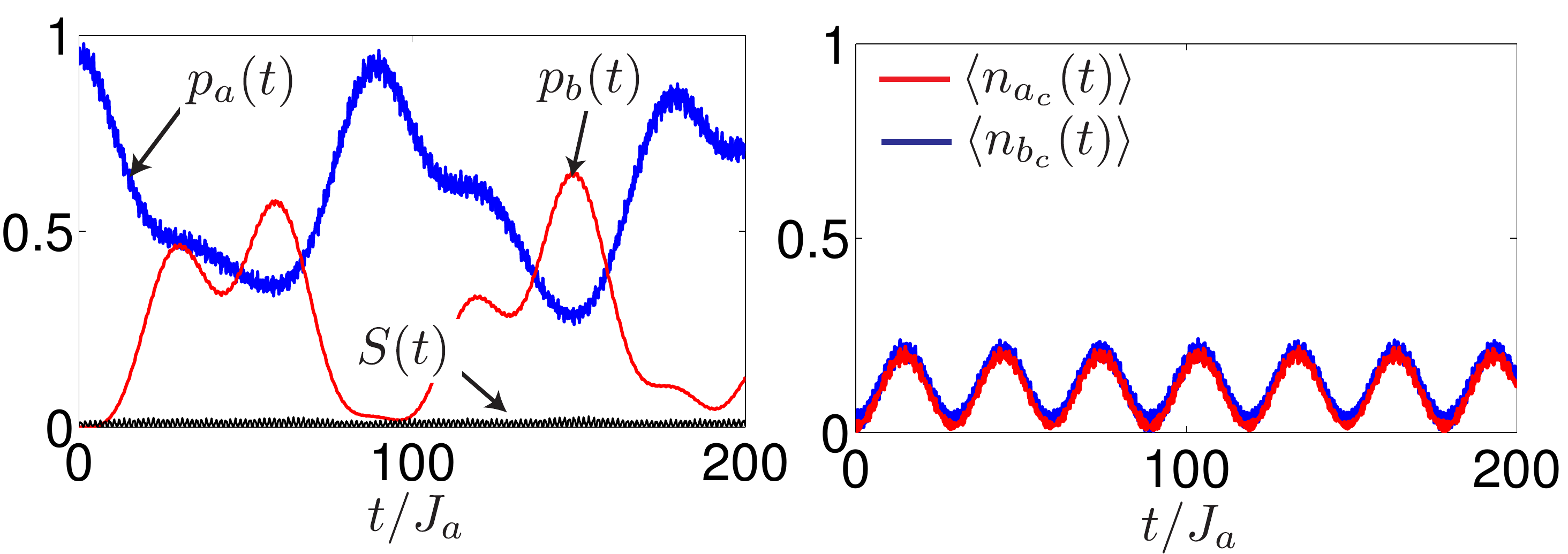}
\caption{Real time evolution for the parameters in the main text and $\delta U = 0 kHz$, $\delta V = 0.02kHz$. a) Time evolution of the expectation values $p_a(t) = \langle n_{a,1}(t) n_{a,2}(t)\rangle$ and $p_b(t) = \langle n_{b,1}(t) n_{b,2}(t)\rangle$ that indicate the pair hopping. Single particle hopping indicated by $S(t) = \sum_{i,j = 1,2} \langle n_{a,i}(t) n_{b,j}(t)\rangle$ is of the order of $10^{-2}$. b) Since the detuning $\delta = 0$, the  occupation of the intermediate sites $\langle n_{a_c}\rangle, \langle n_{b_c}\rangle$, is relatively large large.\label{fig:intermediate}}
\end{figure}

\begin{figure}[t]
\includegraphics[ width = 0.98 \columnwidth]{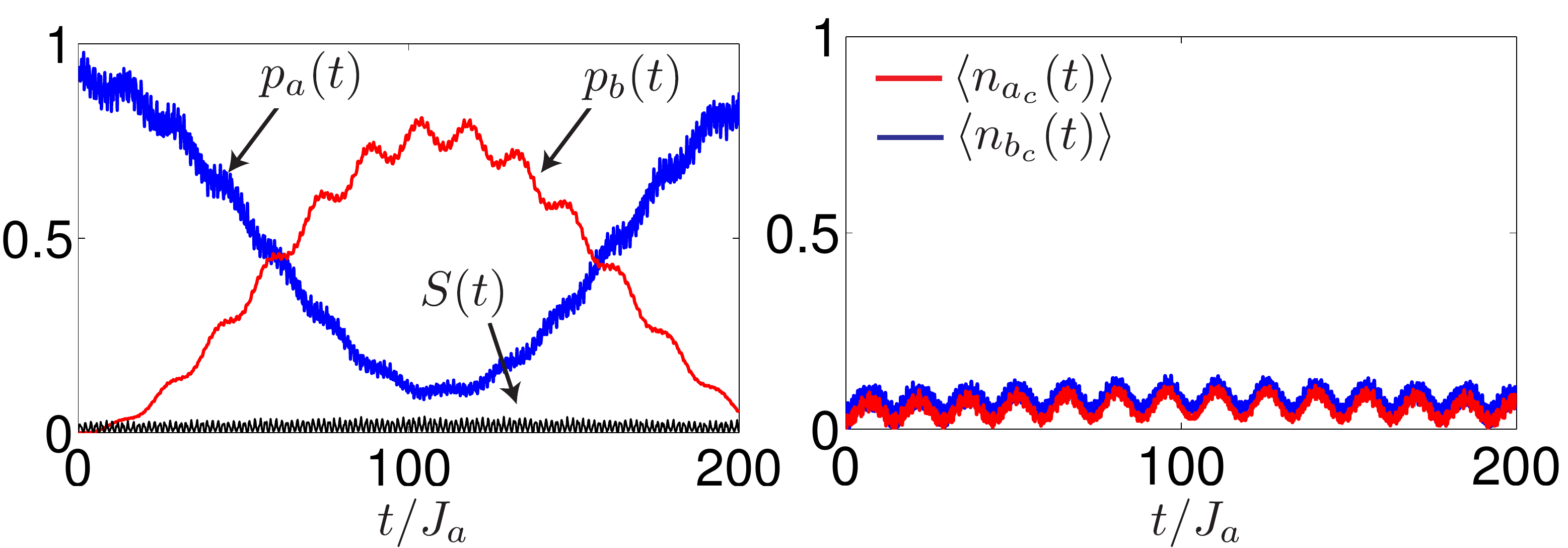}
\caption{Real time evolution for the parameters in  text and $\delta U = 0.5 kHz$, $\delta V = 0.02kHz$. a) Time evolution of the expectation values $p_a(t) = \langle n_{a,1}(t) n_{a,2}(t)\rangle$ and $p_b(t) = \langle n_{b,1}(t) n_{b,2}(t)\rangle$ that indicate the pair hopping. Single particle hopping indicated by $S(t) = \sum_{i,j = 1,2} \langle n_{a,i}(t) n_{b,j}(t)\rangle$ is of the order of $10^{-2}$. b) Occupation of the intermediate sites, $\langle n_{a_c}\rangle, \langle n_{b_c}\rangle$ as a function of time. The detuning $\delta  \neq 0$ leads to relatively small occupation of the intermediate site. \label{fig:optimized}}
\end{figure}

\subsubsection{Suppression of single particle hopping}
We carry out a  numerical analysis that proves the suppression of single particle hopping in model I using the parameters defined above. As a figure of merit, we take the expectation value $S(t) = \sum_{i,j = 1,2} \langle n_{a,i}(t) n_{b,j}(t)\rangle$ that indicates if we find, for some time $t$, one particle on the upper and the other particle on the lower wire. We find, that for $\delta U$ and $\delta V$ on the order of a few hundred Hertz the expectation of $S(t)$ is of the order of $10^{-2}$. 

 \subsubsection{Creation of pair hopping}
 Next, we present a numerical proof that we can engineer pair hopping in our setup. As a figure of merit, we take the expectations $p_a(t) = \langle n_{a,1}(t) n_{a,2}(t)\rangle$ and $p_b(t) = \langle n_{b,1}(t) n_{b,2}(t)\rangle$ indicating that the two particles are both on the same wire. If pair hopping is possible, then $p_a(t)$ should decrease with time, while $p_b(t)$ increases. In Fig.~\ref{fig:intermediate} we present results for $ \delta U = 0$, $\delta V = 0.02kHz$. We see that the occupation of the intermediate site is relatively large. This problem can be overcome by introducing a detuning $\delta U \neq 0$. As shown in Fig. \ref{fig:optimized}, already a small detuning of $\delta U = 0.3 kHz$, $\delta V = 0.02kHz$ significantly reduces the occupation of the intermediate sites. Further, from the width of $p_a(t)$ and $p_b(t)$ we can deduce that $W \sim 1/100 kHz = 10 Hz$ is of the order of the hopping $t_a$.

\subsubsection{Check of the perturbation theory}
Now, we investigate the validity of our perturbative approach. First, we give up the space dependence of the Rabi frequencies and take $\Omega_j = \Omega_{j+1}$. We expect that the pair hopping disappears, and that there is no hopping between the two wires. This assumption is confirmed by our numerical analysis: The expectation $p_a(t) = 1$ for all times, while $p_b(t) = 0$. Further, $S(t)$ is of the order of $10^{-2}$.

\begin{figure}[t]
\includegraphics[ width = 0.95 \columnwidth]{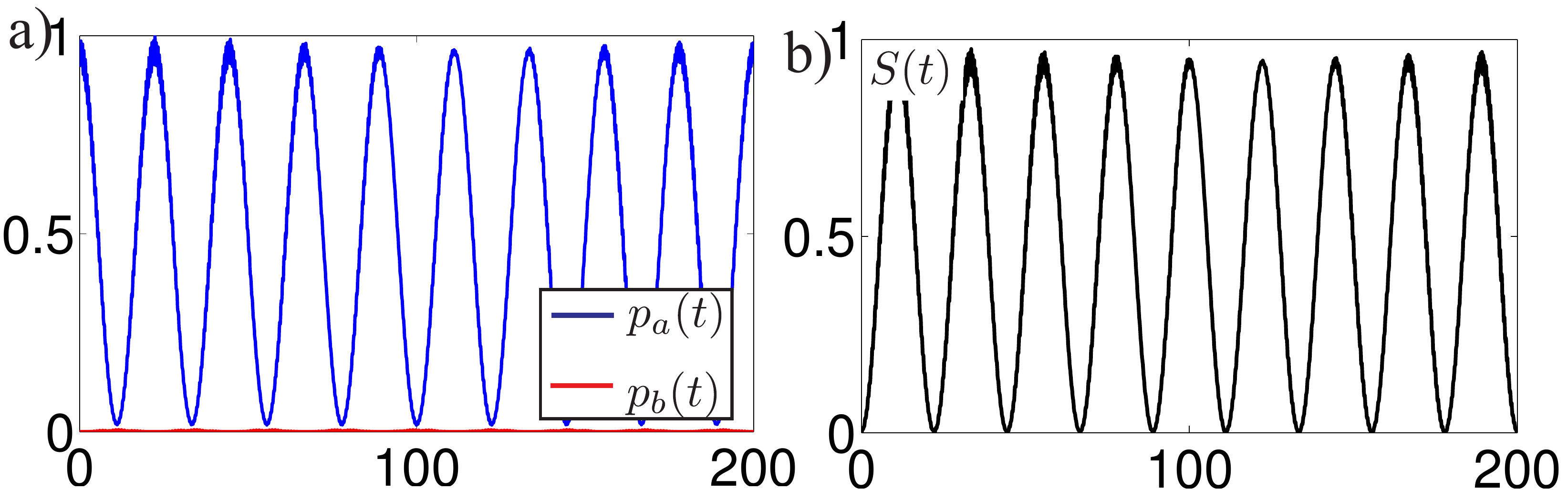}
\caption{Real time evolution for the parameters in the main text and single particle resonance, $\Delta = \hbar \omega_1$. a) Time evolution of the expectation values $p_a(t) = \langle n_{a,1}(t) n_{a,2}(t)\rangle$ and $p_b(t) = \langle n_{b,1}(t) n_{b,2}(t)\rangle$ that indicate a vanishing pair hopping. b) Single particle hopping indicated by $S(t)$ is large. \label{fig:singleparticle}}
\end{figure}

Finally, we consider the case where we tune the parameters in our model such that they are resonant to single particle hopping, i.e.  we take $\Delta = \hbar \omega_1$. We find that $p_b(t) = 0$ for all times in agreement with a vanishing pair hopping (see Fig.~\ref{fig:singleparticle}a)). The results for the single particle hopping, indicated by a non-vanishing values of $S(t)$ are depicted in Fig. \ref{fig:singleparticle}b). 

\section{Final remarks}

Let us conclude with some final remarks on the realization of the pair hopping
Hamiltonian. As discussed in the previous sections, pair tunnelings of order
$W/t\simeq1$ can be achieved when considering tunneling rates $t/h\simeq10-20$
Hz (as currently employed in ultracold fermion experiments~\cite{sugawa2011})
for Rabi frequencies of order $\Omega\simeq2\pi\times6$ kHz. Employing Raman
beams with larger Rabi frequencies would lead to even better energy scales
(although sufficiently strong interactions would be needed as well). An
alternative route toward the observation of Majorana edge states would be to
consider the plaquette Hamiltonians without tracing out intermediate states
with pair of particles in the intermediate sites, that is, obtaining
Hamiltonian of the form in Eq.~\eqref{H2eff}, where the corresponding
couplings $W_{a},W_{b}$ may be much larger that $W$, as they are derived in
second order perturbation theory. The numerical results confirm this
enhancement (in the specific case, by a factor of $\simeq10$). The same
considerations apply to model II. The corresponding many-body scenario would
then be very similar to the one described in Ref.~\onlinecite{fidkowski2011bib},
where the $a$ and $b$-wire are couple to an additional 1D $s$-wave superfluid
with a large spin gap (described by the dynamics in the intermediate sites).

Minimal instances of the present scheme can be validated on a single
plaquette, by measuring the relative parity of a certain state under the
evolution of the effective Hamiltonian by using band-mapping or
\textit{in-situ} imaging techniques~\cite{aidelsburger2011}. Moreover, a
generalization to multi-leg ladders and 2D setups is indeed conceivable under
the same assumptions described in the previous sections.

\end{thebibliography}


\begin{thebibliography}{widest-label}
\bibitem {note1}The parity $P_{1}$ is equal to $(-1)^{(N_{a}-N_{b})/2}$ up to
a constant depending on $N$, and, therefore, the parity $P_{1}$ characterizes
the antisymmetric sector of the theory.
\end{thebibliography}
\end{document}